\documentclass[aps,physrev,preprint,groupedaddress,showkeys]{revtex4-2}

\usepackage{graphicx,changepage}
\usepackage{hyperref}
\usepackage{xcolor}
\usepackage{amsmath}
\usepackage{mathptmx}
\usepackage{etoolbox}
\usepackage{rotating}
\usepackage{booktabs}
\usepackage{ragged2e}
\usepackage{multirow}
\usepackage{gensymb}
\usepackage{physics}

\begin{document}

\title{Anisotropy-driven interfacial magnetism in Ru-deficient SrRuO$_3$ thin films}

\author{V. A. de Oliveira Lima}
\email[E-mail address: ]{v.de.oliveira.lima@fz-juelich.de}
\affiliation{Jülich Centre for Neutron Science for Quantum Materials and Collective Phenomena (JCNS-2), Forschungszentrum Jülich GmbH, 52425 Jülich, Germany}
\affiliation{RWTH Aachen, Faculty of Mathematics, Computer Science and Natural Sciences, 52074 Aachen, Germany}

\author{S. Nandi, T. Brückel}
\affiliation{Jülich Centre for Neutron Science for Quantum Materials and Collective Phenomena (JCNS-2), Forschungszentrum Jülich GmbH, 52425 Jülich, Germany}
\affiliation{RWTH Aachen, Faculty of Mathematics, Computer Science and Natural Sciences, 52074 Aachen, Germany}

\author{M. I. Faley}
\affiliation{Ernst Ruska-Centre for Microscopy and Spectroscopy with Electrons (ER-C-1), Forschungszentrum Jülich GmbH, 52425 Jülich, Germany}

\author{O. Concepción}
\affiliation{Peter Grünberg Institute for Semiconductor Nanoelectronics (PGI-9), Forschungszentrum Jülich GmbH, 52425 Jülich, Germany}

\author{A. Qdemat, H. Ambaye, V. Lauter}
\affiliation{Neutron Scattering Division, Oak Ridge National Laboratory, Oak Ridge, Tennessee  37831, United States of America}

\author{M. Radovic}
\affiliation{PSI Center for Photon Science, Paul Scherrer Institut, CH-5232 Villigen PSI, Switzerland}

\author{A. Singh, E. Kentzinger, C. Bednarski-Meinke}
\affiliation{Jülich Centre for Neutron Science for Quantum Materials and Collective Phenomena (JCNS-2), Forschungszentrum Jülich GmbH, 52425 Jülich, Germany}

\date{\today}

\begin{abstract}
While stoichiometric SrRuO$_3$ (SRO) is a metallic itinerant ferromagnet with relatively homogeneous magnetization, Ru deficiency provides a powerful route to alter its electronic transport and depth-dependent magnetic properties. Ru-deficient SRO thin films grown by radio-frequency high oxygen pressure sputtering were investigated using a combination of X-ray reflectivity, polarized neutron reflectometry, off-specular neutron scattering, scanning transmission electron microscopy with energy-dispersive X-ray spectroscopy, electrical transport, and magnetometry. Structural and compositional analyses reveal that Ru deficiency is intrinsic to the films, with an enhanced deficiency at the interfaces. As a result, coherent electronic transport is suppressed and the saturation magnetization is reduced, while the Curie temperature remains largely unaffected, placing Ru-deficient SRO in a regime consistent with ferromagnetic insulator–like behavior. Depth- and lateral-resolved magnetic measurements further show that the interfacial regions remain ferromagnetic but exhibit enhanced perpendicular magnetic anisotropy, which constrains the local magnetization to remain predominantly out-of-plane and strongly reduces its in-plane projection. Our results establish Ru deficiency as a key control parameter governing transport, magnetization, and anisotropy in SRO thin films and highlight defect and interface engineering as powerful routes to tailor interfacial magnetism in correlated oxide heterostructures.
\end{abstract}

\maketitle
\section{Introduction}

SrRuO$_3$ (SRO) is a complex transition metal oxide that has been used in ferroelectric tunnel junctions and as an electrode material due to its low room-temperature resistivity and high chemical and thermal stability \cite{cuoco2022materials}. Beyond these functional properties, SRO exhibits a wide range of rich and intriguing physical phenomena arising from strong coupling between charge, spin, and orbital degrees of freedom and the periodic crystal lattice. These interactions place SRO at the frontier of materials research for both conventional electronics and emerging quantum technologies \cite{gu2022overview, wakabayashi2021single}.

SRO stands out as a rare example of a 4$d$ transition metal oxide that simultaneously exhibits itinerant ferromagnetism and metallic conductivity below its Curie temperature ($T_{\text{Curie}}$) of 160 K. More recently, it has attracted renewed interest as a platform for studying correlated and topological phenomena \cite{oliveira2025anomalous,takiguchi2020quantum,huang2020detection}. For instance, SRO displays strain-tunable magnetic anisotropy with narrow domain walls, strong spin–orbit coupling, and unconventional transport behavior, including anomalous and topological Hall effects, linked to nontrivial Berry curvature \cite{zheng2025manipulating,takiguchi2020quantum,chen2013weyl,marshall1999lorentz}. Moreover, recent theoretical predictions suggest that SRO may host topological semimetal phases, such as Weyl points and Dirac nodal lines near the Fermi level \cite{sohn2021sign}. Its physical properties are remarkably sensitive to epitaxial strain, dimensional confinement, interfacial effects, and film stoichiometry \cite{wakabayashi2021wide, xia2009critical,toyota2005thickness}.

Among these parameters, Ru deficiency plays a particularly important role in SRO thin films. Rather than acting as a simple source of disorder, Ru vacancies strongly modify the electronic bandwidth, magnetic exchange interactions, and magnetic anisotropy \cite{zhang2023tuning,wakabayashi2021structural,takiguchi2020quantum,koster2012structure}. As a result, Ru-deficient SRO films can deviate substantially from the metallic itinerant ferromagnetic ground state of stoichiometric SRO, particularly at interfaces where stoichiometric disorder and anisotropy gradients are enhanced \cite{wakabayashi2021structural,mlynarczyk2024physicochemical}. Wakabayashi et al. conducted a comparative study of the structural and electronic transport properties of stoichiometric and Ru-deficient SRO thin films grown by molecular beam epitaxy (MBE) \cite{wakabayashi2021structural}. They observed two primary types of structural disorder in Ru-deficient films (Ru vacancies within the lattice and interface-driven disorders) which are thickness-dependent. These structural features were correlated with an enhanced magnetic moment, a reduction in $T_{\text{Curie}}$, and the suppression of the magnetic Weyl semimetal state. 

In contrast, Mlynarczyk et al. investigated SRO films grown by direct current high oxygen pressure sputtering (DC-HOPS) and found that the Ru deficiency, approximately 14$\%$, was localized at both film interfaces (i.e. at the surface and at the film–substrate interface), while the bulk of the film remained stoichiometric \cite{mlynarczyk2007surface}. Despite the presence of Ru deficiency, the crystallinity of such films remains high, as evidenced by well-defined finite-thickness (Laue) oscillations in X-ray diffraction (XRD) measurements. The only notable difference between stoichiometric and Ru-deficient samples is a slight out-of-plane lattice expansion \cite{siemons2007dependence}. These observations highlight the complex nature of disorder in SRO thin films: even in the absence of major structural degradation, subtle modifications in stoichiometry can lead to significant changes in magnetic and electronic behavior. 

In this work, we investigate Ru-deficient SRO thin films grown on Nb-doped and undoped SrTiO$_3$ substrates by radio-frequency high oxygen pressure sputtering (RF-HOPS). Similar to the works of Wakabayashi et al. and Mlynarczyk et al., our films also exhibit signatures of Ru deficiency and interfacial inhomogeneity. While previous studies on such films have largely focused on transport properties and volume-averaged magnetic measurements, the impact of Ru deficiency on the depth-dependent magnetic structure, particularly at buried interfaces, remains poorly understood. Here, we address this issue by combining X-ray reflectivity and polarized neutron reflectometry to directly probe the nuclear and magnetic scattering length density profiles with nanometer resolution. Our approach provides new insights into how stoichiometry variations and interface quality affect the magnetic properties of Ru-deficient SRO thin films and is broadly relevant for defect-engineered control of interfacial magnetism in correlated oxide heterostructures.

\section{Experimental details}

Single-crystal Nb-doped (Nb:STO, 0.05 wt$\%$ Nb, $\approx$ 0.1 at$\%$) and undoped SrTiO$_3$ (STO) (001) substrates, supplied by Shinkosha, with miscut angles between 0.05$^{\circ}$ and 0.1$^{\circ}$, were used for thin film growth. While undoped STO is a standard substrate for high-quality SRO thin films, it undergoes an antiferrodistortive (AFD) structural phase transition at approximately 105 K, which leads to the formation of structural domains and associated strain fields near the surface that can affect neutron scattering experiments and the magnetic response of adjacent thin films \cite{hoppler2008x}. To avoid such effects, polarized neutron reflectometry and off-specular neutron scattering measurements were performed exclusively on films grown on Nb:STO substrates, for which Nb incorporation modifies the AFD transition temperature and domain formation behavior, reducing structural inhomogeneity at low temperatures \cite{zhang2022manipulating, mccalla2016unified}. Prior to deposition, all substrates were chemically etched using buffered HF and subsequently annealed at 950 $^{\circ}$C for 2 h in air to obtain a well-defined TiO$_2$-terminated surface \cite{sanchez2014tailored}.

The SRO thin films were deposited from a stoichiometric SRO target using RF-HOPS \cite{poppe1992low,jia2002lattice,faley2011epitaxial}. 
To guarantee a high-quality thin film growth and an environment free of contaminants, the SRO target is pre-sputtered, and the deposition chamber is pumped for 24 h before film deposition. The SRO thin film was deposited at $T_{\text{dep}}$ = 785 $\degree$C using a target-substrate distance (D$_{\text{TS}}$) of 2.5 cm, oxygen pressure (P$_{\text{O$_2$}}$) of 1.5 mbar, and RF plasma generated using a forward power (FWDP) of 100 W. The SRO growth rate at such growth parameters is about 125 Å/h \cite{oliveira2025anomalous}. 

The sample morphology and local roughness were analyzed using Atomic Force Microscopy (AFM) in intermittent contact mode, employing an Agilent 5400 AFM/SPM system equipped with Mikromash HQ:NSC15 cantilevers. Epitaxial quality, layer thickness, and total surface roughness were examined through X-ray diffraction (XRD), X-ray reflectivity (XRR), reciprocal space mapping (RSM), and $\phi$-scan measurements, conducted on a Rigaku SmartLab diffractometer. The crystal structure and the chemical composition across the film, with particular focus on the top and bottom interfaces, were investigated by high-angle annular dark-field (HAADF) scanning transmission electron microscopy coupled with energy-dispersive X-ray spectroscopy (STEM-EDS) using TFS Spectra 300 STEM and FEI Titan G2 80-200 ChemiSTEM. 

Magnetic hysteresis and field-cooled magnetization measurements were performed as a function of temperature and applied magnetic field using a Superconducting Quantum Interference Device (SQUID) Quantum Design Magnetic Properties Measurement System (MPMS-XL). The temperature dependence of the electrical transport properties was investigated by standard four-point probe resistivity measurements using a Quantum Design Physical Property Measurement System (PPMS).

Polarized neutron reflectivity (PNR) and off-specular neutron scattering (OSS) were measured on the Magnetism Reflectometer, MAGREF \cite{lauter2009highlights}, at BL-4A of the Spallation Neutron Source, Oak Ridge National Laboratory, using a 10$\times$10 mm$^2$ sample. Reflectivity was measured up to wave vector transfer ($\vb Q_\text{z}$) of 0.12 Å$^{-1}$ at 100 K, 80 K, and 5 K, following field cooling in a 4.8 T in-plane magnetic field (applied along the [100] crystallographic direction). The neutron spin polarization was set either parallel (R+) or antiparallel (R--) to the field direction, and was applied throughout the measurements. 

\section{Results and discussion}

\begin{figure*}
    \includegraphics[width=0.9\linewidth]{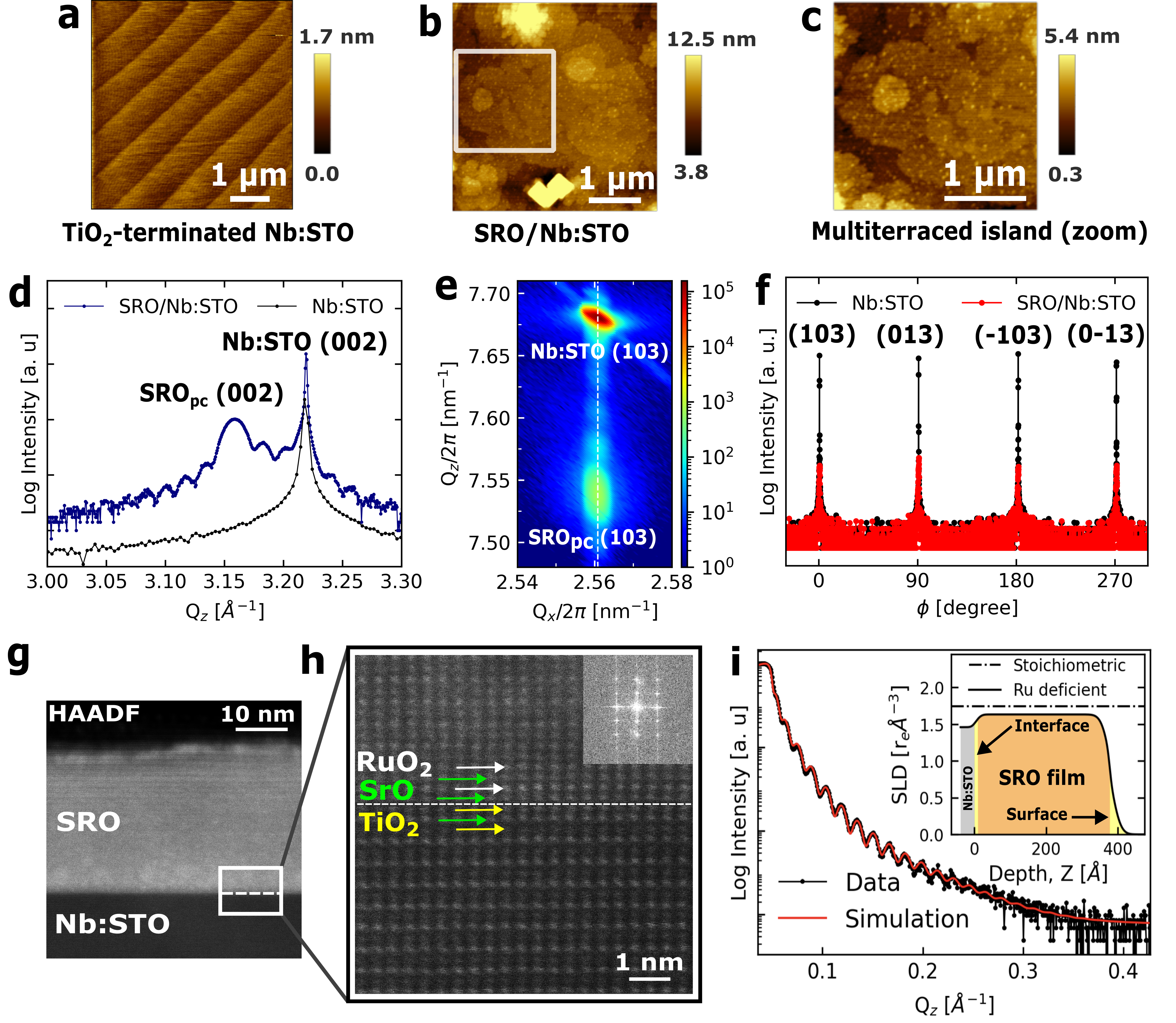}
    \caption{Atomic force microscopy micrographs of (a) the Nb:STO substrate and (b) the SRO thin film grown on Nb:STO. (c) Magnified view of the highlighted region in (b). The magnified image highlights the concentric multiterraced structure of an individual island. (d) X-ray diffraction pattern showing clear Laue oscillations, indicating the high crystalline quality of the SRO film. (e) Reciprocal space map and (f) $\phi$-scans measured around the (103) crystallographic reflection of the SRO thin film and Nb:STO substrate. (g) Cross-sectional high-angle annular dark-field scanning transmission electron microscopy (HAADF-STEM) image of the epitaxial SRO film on the Nb:STO substrate. (h) Atomic-resolution image of the interface, revealing a sharp transition with alternating RuO$_2$/SrO/TiO$_2$ planes; the white dashed line marks the position of the SRO/Nb:STO interface. The inset shows the fast Fourier transform of the atomic-resolution image. (i) X-ray reflectivity data (black symbols) and corresponding GenX simulations (red line) based on a structural model that includes reduced scattering length density (SLD) layers at both the surface and the bottom interface (SRO/Nb:STO). The inset shows the real part of the extracted SLD depth profile. The dash-dotted horizontal line indicates the nominal X-ray SLD expected for stoichiometric SRO, shown as a reference.}
    \label{fig:AFM_XRD}
\end{figure*}

Unless otherwise stated, the structural and magnetic results presented in this section refer to Ru-deficient SRO films grown on Nb:STO substrates, which were selected for neutron scattering experiments to avoid substrate-related structural artifacts. The Nb:STO substrate exhibits the characteristic terraced morphology associated with TiO$_2$ surface termination [Fig. \ref{fig:AFM_XRD}(a)]. Its root mean square (RMS) local roughness calculated from 5 $\times$ 5 $\mu$m$^2$ AFM scans is approximately 8 Å $\pm$ 2 Å. 

The morphology of the deposited 375 Å (determined from the calibrated SRO growth rate) SRO thin film [Fig. \ref{fig:AFM_XRD}(b)] reveals evidence of an Stranski–Krastanov  growth mode (layer-plus-island growth), as indicated by the presence of multiterraced islands. Such growth behavior may be influenced by several factors, including the substrate miscut angle \cite{sanchez2014tailored,rao1997growth}, the terraces alignment in relation to the [010] substrate direction (also called angle of miscut direction) \cite{gan1997control,wang2020magnetic,sanchez2014tailored}, and the substrate surface energy \cite{iles2010systematic}. Figure \ref{fig:AFM_XRD}(c) shows a magnified view of the region marked in Fig. \ref{fig:AFM_XRD}(b), emphasizing the contours of a multiterraced island. The presence of multiple concentric terraces surrounding an individual island nucleus suggests that secondary nucleation occurs on top of preexisting islands. Importantly, the island morphology appears largely isotropic in the plane of the film, indicating that the directional information associated with the substrate step edges is not preserved during growth. As a consequence, any step-edge–induced anisotropy is strongly reduced in the SRO layer. This isotropic island growth and relatively high roughness are consistent with the coexistence of multiple in-plane crystallographic domain variants. The RMS local roughness of the SRO film, calculated over an area of 2.5 $\times$ 2.5 $\mu$m$^2$, was 37 Å $\pm$ 14 Å.

The high crystalline quality of the SRO thin film is evidenced by the presence of Laue oscillations in the XRD pattern [Fig. \ref{fig:AFM_XRD}(d)]. The film shows epitaxial growth, with an in-plane lattice parameter that is compressed and aligned with that of the substrate, as confirmed by the RSM and $\phi$-scans performed around the pseudocubic (103) crystallographic reflection of the SRO thin film and Nb:STO substrate [Fig. \ref{fig:AFM_XRD}(e–f)]. The slightly larger out-of-plane lattice parameter obtained from the RSM (3.98 Å ± 0.02 Å), which is larger than for stoichiometric SRO films (3.949 Å), provides a first indication of Ru deficiency in the film \cite{wakabayashi2021structural,siemons2007dependence}. The in-plane crystallographic structure was further examined by $\phi$-scan measurements around the SRO$_{\text{pc}}$ (103) reflection [Fig. 1(e–f)]. The $\phi$-scan reveals four well-defined peaks separated by 90$^{\circ}$, evidencing the coexistence of two in-plane orthorhombic (110) domain variants. The comparable peak intensities indicate similar domain populations, consistent with previous reports on SRO films grown on exact or low-miscut TiO$_2$-terminated STO substrates \cite{gan1997control,jiang1998domain}. Such multidomain growth is commonly observed when no dominant step-edge anisotropy is imposed by the substrate, in agreement with the isotropic island morphology observed by AFM.

Cross-sectional high-angle annular dark-field scanning transmission electron microscopy (HAADF-STEM) reveals a fully epitaxial SRO film grown on Nb:STO with a coherent and atomically sharp interface, confirming the high crystalline quality of the heterostructure [Fig. \ref{fig:AFM_XRD}(g)]. The atomic-resolution image in Fig. \ref{fig:AFM_XRD}(h) clearly shows the alternating RuO$_2$, SrO, and TiO$_2$ planes, indicative of an ordered perovskite stacking across the interface. The white dashed line marks the position of the SRO/Nb:STO interface. The inset in Fig. \ref{fig:AFM_XRD}(h) shows the fast Fourier transform (FFT) of the atomic-resolution image, confirming the high degree of crystallinity and epitaxial coherence. It is worth noting that, at the Ru deficiency levels inferred from compositional analyses, the perovskite stacking remains intact and structural signatures of Ru vacancies are not directly discernible in atomic-resolution HAADF-STEM images, which primarily probe the local structural order.

To quantitatively determine the total film thickness ($t$) and interfacial roughness ($\sigma$), XRR data were analyzed using the GenX software \cite{glavic2022genx}. Several structural models, based on stacked block-like layers on a substrate, were implemented to represent the layered architecture of the film. A schematic overview of these models, along with their key parameters, is provided in the Supplementary Material (SM, section \ref{section:sup_s1}). While all tested models successfully reproduced the overall periodicity of the XRR oscillations (SM, section \ref{section:sup_s2}), primarily sensitive to the film thickness and interfaces, only one model (model 4) adequately captures the finer features associated with interfacial roughness. This model includes reduced scattering length density (SLD) interfaces at both the top (surface) and bottom (SRO-Nb:STO) film boundaries. The experimental XRR pattern and the corresponding simulation are shown in Fig. \ref{fig:AFM_XRD}(i). 

The total film thickness is $t = 382 \pm 3$ Å, corresponding to the sum of all sublayer thicknesses and consistent with the expected thickness based on the calibrated SRO growth rate. In contrast to the total geometrical thickness obtained from XRR, the finite-thickness (Laue) oscillations observed around the SRO$_{pc}$ (002) reflection yield a structurally coherent crystalline thickness of $t_{\text{coh}}$ = 255 Å $\pm$ 6 Å. The smaller coherent thickness compared to the XRR-derived value reflects the distinct physical quantities accessed by the two techniques: while XRR is sensitive to the full electron density profile, including chemically and structurally modified surface and SRO/Nb:STO interface regions, Laue oscillations arise only from the coherently ordered fraction of the film. In Ru-deficient SRO, vacancy-induced disorder and strain inhomogeneities are expected to reduce the out-of-plane coherence length, consistent with the vacancy-driven structural degradation discussed in \cite{wakabayashi2021structural}. 

The roughness parameters extracted for the substrate/film interface, internal film interfaces, and the film surface were $\sigma = 10$ Å, 5 Å, and 21 Å, respectively.  Assuming uncorrelated interfaces, these values yield an average effective interfacial roughness of $\sigma_{\mathrm{eff}} \approx 13.7 \pm 2$ Å, estimated by an RMS combination of the individual roughness parameters. This assumption is physically justified by the different origins of roughness at the substrate and film interfaces: substrate roughness is dominated by step–terrace structures associated with the miscut, whereas film roughness arises from growth-induced morphology and island coalescence, leading to largely independent height fluctuations.

The thicknesses of the surface and SRO–STO interface regions are estimated to be in the ranges of 4–10 Å and 10–30 Å, respectively, in good agreement with values reported by Mlynarczyk et al. \cite{mlynarczyk2024physicochemical,mlynarczyk2007surface}. In those studies, Ru deficiency was found to be concentrated at both the top and bottom interfaces of the film, leading to Sr enrichment in these regions, while the bulk of the film remained stoichiometric \cite{mlynarczyk2007surface}. Here, our XRR results provide a quantitative structural signature of this compositional variation by revealing a reduced X-ray SLD at the interfaces. Moreover, Ru deficiency is not confined to the interfaces but extends throughout the entire film thickness, as evidenced by the reduced SLD in the central region (1.633 ± 0.024 r$_\mathrm{e}$Å$^{-3}$) compared to the theoretical X-ray SLD of stoichiometric SRO (1.746 r$_\mathrm{e}$Å$^{-3}$), indicated as a dash-dotted horizontal line [see inset of Fig. \ref{fig:AFM_XRD}(i)]. Assuming constant atomic density and full oxygen stoichiometry, the reduced X-ray SLD corresponds to an effective Ru deficiency of 24$\%$, providing an upper bound for the Ru vacancy concentration (SM, section \ref{section:sup_S3}).

\begin{figure*}   
    \includegraphics[width=\linewidth]{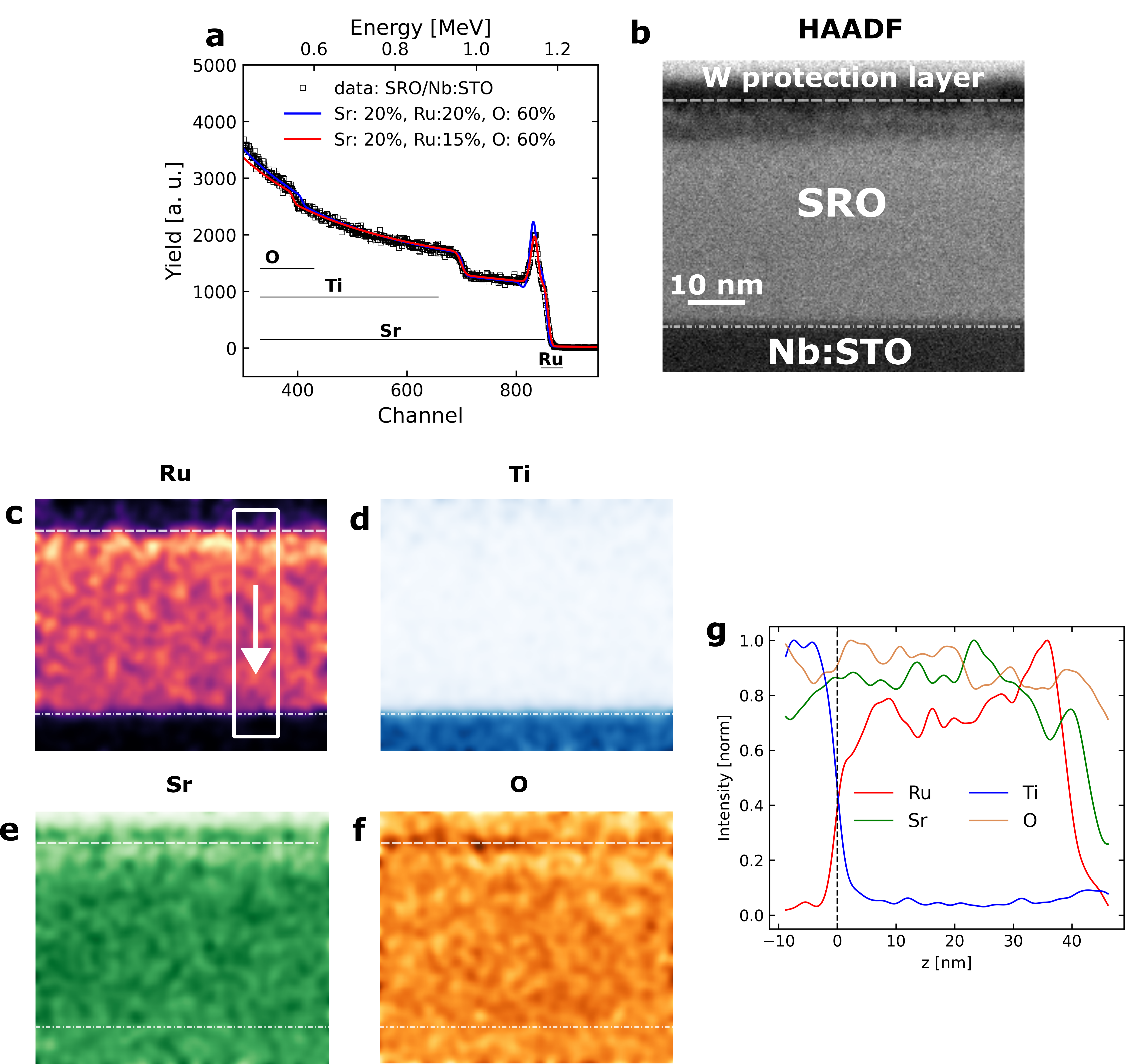}
    \caption{(a) Rutherford backscattering spectroscopy spectrum of the SRO/Nb:STO thin film together with RUMP simulations assuming stoichiometric SrRuO$_3$ (blue line) and a Ru-deficient composition (SrRu$_{0.75}$O$_3$, red line). (b) Cross-sectional HAADF-STEM image of the region used for compositional mapping. (c–f) Corresponding STEM-EDS elemental maps of Ru, Ti, Sr, and O, respectively. The dashed lines in panels (c–f) are guides to the eye marking the approximate positions of the film surface and the buried SRO/Nb:STO interface. (g) Laterally integrated and normalized STEM-EDS intensity profiles extracted from the region indicated in panel (c), plotted as a function of depth $z$, where $z = 0$ corresponds to the SRO/Nb:STO interface.}
    \label{fig:TEM}
\end{figure*}

This interpretation is further supported by Rutherford backscattering spectroscopy (RBS) measurements. Figure \ref{fig:TEM}(a) shows the RBS spectrum of the SRO/Nb:STO film together with RUMP simulations using two representative compositions: stoichiometric SrRuO$_3$ and a Ru-deficient model (SrRu$_{0.75}$O$_3$). The Ru content predominantly affects the feature around $\sim$1.1 MeV and the high-energy shoulder of the Ru signal, for which the Ru-deficient model provides a significantly improved agreement with the experimental spectrum. This indicates that the film is globally Ru-poor, with an effective Ru deficiency on the order of $\sim$20–25$\%$. We note that, as a depth-integrated and heavy-element-weighted technique under the present measurement conditions, RBS primarily constrains the overall cation stoichiometry and is comparatively less sensitive to interfacial composition gradients and oxygen non-stoichiometry.

To obtain spatially resolved compositional information and directly probe interfacial chemical gradients, STEM-EDS was performed in the cross-sectional region [Fig. \ref{fig:TEM}(b)]. Elemental maps for Ru, Ti, Sr, and O are presented in Figs. \ref{fig:TEM}(c–f). Laterally integrated STEM-EDS intensity profiles are shown in Fig. \ref{fig:TEM}(g) to facilitate a direct comparison of the depth-dependent elemental distributions across the sample. The profiles were extracted from the region indicated in Fig. \ref{fig:TEM}(c) and are plotted as a function of depth using the same convention adopted in the X-ray and neutron reflectometry analyses, with the SRO/Nb:STO interface defined as $z = 0$. 

The Ru map reveals pronounced compositional gradients, with reduced Ru signal near both the film surface and the buried SRO/Nb:STO interface, qualitatively consistent with the reduced SLD layers inferred from XRR. The Ru depletion appears stronger at the buried interface than near the surface, suggesting that Ru deficiency is enhanced at the substrate side. In addition, a slightly Ru-enhanced region is observed just below the surface, which may indicate a non-uniform Ru distribution across the film thickness (e.g., local Ru accumulation during growth), although quantification is limited by the spatial resolution and thickness integration of the EDS measurement.

The Ti map shows no clear evidence of Ti diffusion into the SRO layer (within the sensitivity of STEM-EDS), supporting chemically well-defined interfaces. The Sr and O maps exhibit comparatively weaker variations; regions of reduced Ru signal tend to coincide with a slight relative Sr enhancement, consistent with cation off-stoichiometry. A fully quantitative description would additionally require assessing possible oxygen non-stoichiometry, which is not robustly constrained by the present RBS/EDS data. The combined RBS and STEM-EDS results corroborate the XRR-based structural model and support a picture of Ru deficiency that is enhanced at interfaces and present throughout the film.

The high volatility of Ru can lead to the formation of distinct types of Ru-vacancy-related disorder, as identified by Wakabayashi et al. \cite{wakabayashi2021structural}. One contribution arises from Ru vacancies distributed throughout the bulk of the film, introducing structural disorder that can reduce the coherent crystalline thickness relative to the total thickness measured by XRR. A second contribution is interface-driven disorder, associated with compositional and structural variations localized at the film–substrate interface, which can influence the interfacial quality and the depth-dependent SLD profile. In this context, XRR provides the quantitative link between the depth-integrated stoichiometry obtained from RBS and the local compositional variations revealed by STEM-EDS, yielding a statistically averaged depth profile of Ru deficiency across the film. This combined structural and compositional picture establishes a consistent physical framework for understanding the properties of Ru-deficient SRO.

\begin{figure*}[hbtp]
    \includegraphics[width=0.85\linewidth]{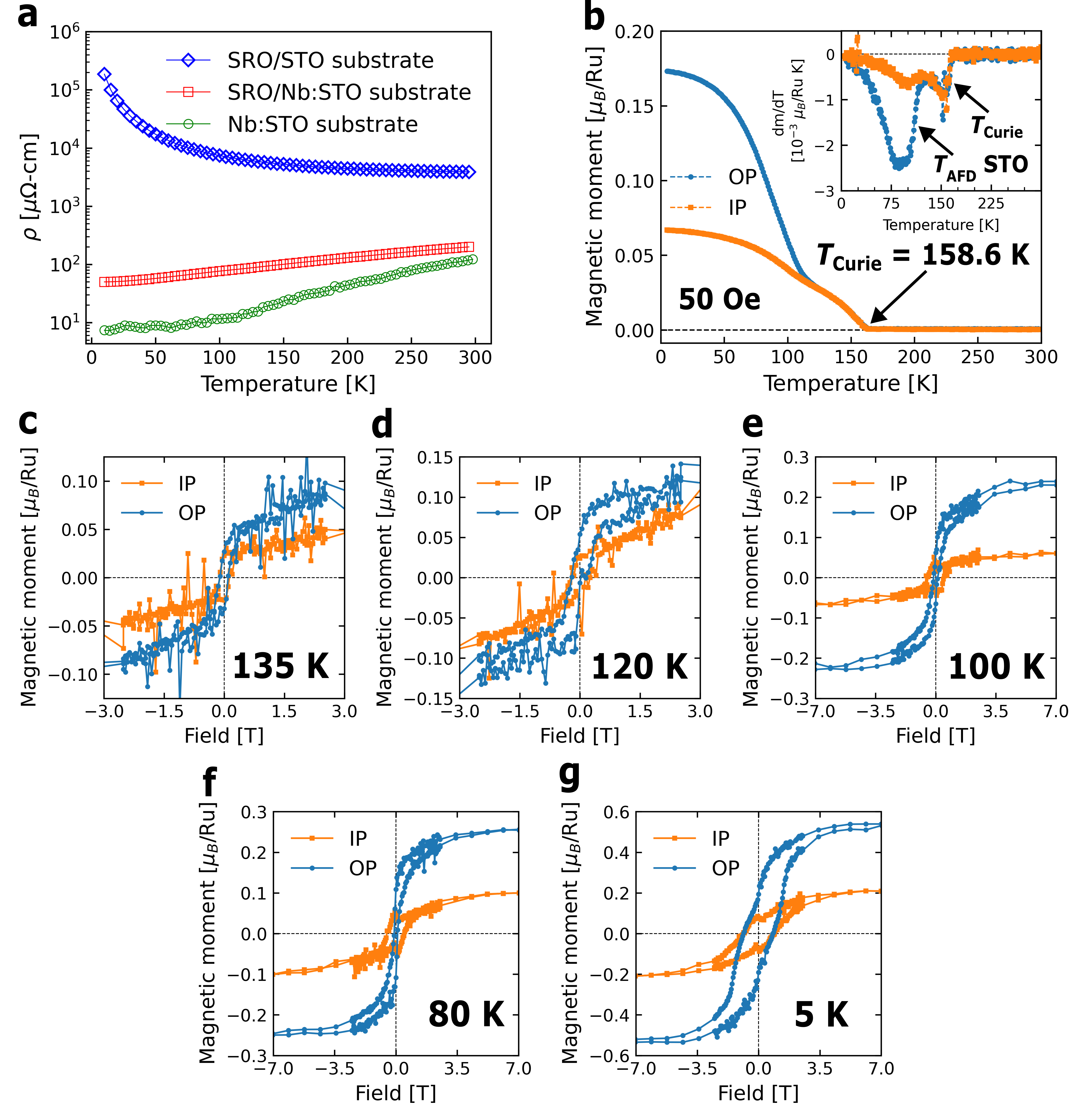}
    \caption{(a) Temperature dependence of the resistivity for SRO thin films grown on undoped STO and Nb-doped STO substrates. (b) Field-cooled magnetization measured under an applied field of 50 Oe, with the field oriented perpendicular (out-of-plane, OP) and parallel (in-plane, IP) to the film surface. The inset shows the first derivative of the magnetization, highlighting the Curie temperature ($T_\text{Curie}$) and the antiferrodistortive transition temperature ($T_\text{AFD}$) of the STO substrate. (c,d) Magnetic hysteresis loops M($H$) measured at 135 K and 120 K, respectively, showing weak perpendicular magnetic anisotropy (PMA) with nearly identical OP and IP responses. (e) M($H$) measured at 100 K, revealing the onset of a clear bifurcation between OP and IP magnetization, indicative of a crossover toward enhanced PMA. (f,g) M($H$) measured at 80 K and 5 K, respectively, displaying a pronounced enhancement of PMA, with easier saturation along the OP direction. For clarity, the hysteresis loops at 135 K and 120 K are shown up to 3 T due to increased noise at higher fields. All M($H$) were performed after field cooling the sample in 5 T.}
    \label{fig:transmag}
\end{figure*}

Electron transport properties are strongly affected by Ru deficiency, as evidenced by the temperature dependence of the resistivity [Fig. \ref{fig:transmag}(a)] for films grown on both Nb-doped and undoped STO substrates. The absence of a discernible kink at the Curie temperature ($T_{\text{Curie}} \approx 160$ K) and the pronounced increase in resistivity at low temperatures indicate semiconducting or insulating behavior within the measured temperature range. The insulating response observed for the film grown on the undoped STO substrate demonstrates that the Ru-deficient SRO layer itself is intrinsically non-metallic, consistent with previous reports on Ru-poor SRO thin films \cite{wakabayashi2021structural}. For the film grown on the Nb-doped substrate, the resistivity curve exhibits an apparent metallic-like temperature dependence. However, this behavior is dominated by a shunting effect, whereby the electrical current preferentially flows through the highly conductive Nb:STO substrate rather than through the SRO layer. As a result, the measured resistivity underestimates the intrinsic resistivity of the Ru-deficient SRO film, as it includes a substantial contribution from the conductive substrate. This decoupling between electronic transport and magnetic order reflects the sensitivity of Ru–O–Ru hopping to vacancy disorder, while the ferromagnetic exchange interaction remains comparatively robust.

The temperature dependence of the magnetization $M(T)$, measured after field cooling the sample in a magnetic field of 50 Oe applied perpendicular (out-of-plane) and parallel (in-plane) to the film surface, is shown in Fig. \ref{fig:transmag}(b). The $T_{\text{Curie}}$ of 158.6 K $\pm$ 0.5 K was extracted from the first derivative of the FC data [inset of Fig. \ref{fig:transmag}(b)]. Complementary magnetization measurements performed on Ru-deficient SRO films grown on undoped STO substrates are provided in the SM (section \ref{section:sup_complement}, Fig. \ref{fig:sup3}). The $T_{\text{Curie}}$ is not strongly affected by the Ru deficiency in SRO films prepared by RF-HOPS, which contrasts with $T_{\text{Curie}}$ values of Ru-poor SRO films prepared by MBE ($\sim$140 K) and pulsed laser deposition ($\sim$143 K), where similar levels of non-stoichiometry have been reported to lower $T_{\text{Curie}}$ \cite{wakabayashi2021structural, yoo2005contribution}. These observations highlight the distinct behavior of SRO films grown by RF-HOPS, suggesting that this technique not only influences the spatial distribution of Ru vacancies but may also mitigate their detrimental effects on magnetic ordering. This combination of non-metallic transport behavior and a robust $T_{\text{Curie}}$ is consistent with a ferromagnetic insulator–like regime induced by Ru-vacancy-related disorder.

In addition to the ferromagnetic transition, a distinct feature emerges in the FC magnetization below $T_{\text{Curie}}$, starting at approximately 117 K and peaking near 87 K. This temperature range coincides with the AFD cubic-to-tetragonal structural transition of the STO substrate, which typically occurs between 110 K and 65 K \cite{hoppler2008x}. Simultaneously, a clear bifurcation between the out-of-plane and in-plane magnetization components develops in Fig. \ref{fig:transmag}(b), indicating an enhancement of perpendicular magnetic anisotropy (PMA) upon cooling. The persistence of this feature suggests that Nb doping does not suppress the AFD transition, but rather modifies its onset temperature and domain configuration \cite{mccalla2016unified,zhang2022manipulating}, potentially influencing the magnetic anisotropy of the adjacent Ru-deficient SRO film.

The temperature evolution of the magnetic anisotropy is further elucidated by the magnetic hysteresis loops $M(H)$ shown in Figs. \ref{fig:transmag}(c–g). All hysteresis loops were recorded after field cooling under an applied magnetic field of 5 T. At 135 K and 120 K [Figs. \ref{fig:transmag}(c,d)], the out-of-plane (OP) and in-plane (IP) $M(H)$ are nearly identical, indicating only weak PMA in this temperature range. Upon further cooling to 100 K [Fig. \ref{fig:transmag}(e)], a clear bifurcation between the OP and IP magnetization emerges, marking a crossover toward enhanced PMA. This crossover temperature coincides with the onset of the OP–IP separation observed in the FC $M(T)$ [Fig. \ref{fig:transmag}(b)]. At lower temperatures, 80 K and 5 K [Figs. \ref{fig:transmag}(f,g)], the anisotropy becomes pronounced, with the OP direction acting as a clear easy axis, characterized by sharp magnetization reversal and easier saturation compared to the IP hard axis. The progressive enhancement of PMA upon cooling suggests that structural and interfacial effects, potentially amplified by the AFD transition of the Nb:STO substrate, may play an important role in stabilizing the magnetic anisotropy in Ru-deficient SRO films.

At the lowest temperatures, additional features appear in the hysteresis loops, most notably small irregularities in the magnetization reversal, particularly at 5 K. Such step-like features are characteristic of multidomain behavior commonly observed in strained SRO thin films \cite{jiang1998domain,gan1997control}. In this regime, multiple crystallographic and magnetic domain variants can coexist, giving rise to non-smooth magnetization reversal processes \cite{wang2020magnetic,zahradnik2020magnetic,marshall1999lorentz}. These multidomain states are consistent with terrace-induced IP variants arising from the low substrate miscut and terrace orientation. 

In stoichiometric SRO, the OP saturation magnetic moment per Ru atom typically reaches about 1.6 $\mu_{B}$ at low temperatures \cite{koster2012structure}. In contrast, the present Ru-deficient film exhibits a significantly reduced effective magnetic moment of approximately 0.53 $\mu_{B}$ at 5 K. This value was obtained by normalizing the SQUID magnetization to the effective magnetic moment per Ru atom following the procedure described in the SM (section \ref{section:sup_conversion}). This reduction evidences the disruption of the Ru–O–Ru exchange interactions that mediate ferromagnetism in SRO \cite{koster2012structure}, as Ru vacancies locally break these exchange paths, introducing magnetic frustration and weakening the double-exchange mechanism responsible for long-range order \cite{wakabayashi2021structural,palai2009observation}. Nevertheless, the persistence of a robust ferromagnetic signal indicates that the vacancy concentration remains below the threshold that would fully suppress magnetic ordering.

The high-field saturation magnetization also shows a clear dependence on the direction of the applied magnetic field, a behavior commonly reported for SRO thin films. Several factors may contribute to this anisotropic response \cite{goossens2021anisotropy}. Ziese et al.  noted that the typical procedure for subtracting the diamagnetic substrate background, which is done by removing the extrapolated high-field slope, may neglect paramagnetic contributions from the film itself, potentially leading to incorrect estimates of the saturation magnetization along different crystallographic directions \cite{ziese2010structural}. Moreover, the anisotropy field in SRO can exceed the maximum applied magnetic field, thereby preventing full saturation magnetization along the magnetic hard axes \cite{koster2012structure}. Structural defects may also contribute to the suppression of full saturation by disrupting magnetic order \cite{zahradnik2020magnetic}. 

The SQUID magnetization curves [Fig. \ref{fig:transmag}] provide values directly proportional to the total volume magnetization of the sample and, as such, do not offer the spatial resolution necessary to probe interface-specific magnetic behavior. In contrast, PNR provides depth-resolved information on both the nuclear and magnetic structure of thin films and buried interfaces \cite{lauter2007neutron,daillant2008x}. The nuclear scattering length density (nSLD) describes the atomic structure, whereas the magnetic scattering length density (mSLD) reflects the IP component of the magnetization as a function of depth \cite{lauter2007neutron,daillant2008x,toperverg2015neutron,lauter2022m,dronskowski2024neutron}. 

To investigate the influence of Ru deficiency on the magnetic depth profile of SRO thin films, PNR measurements were performed at selected temperatures following field cooling in a 4.8 T magnetic field applied parallel to the film surface (IP). During the measurements, the magnetic field remained applied IP and the neutron spin polarization was set parallel (R$+$) or antiparallel (R$-$) to this direction, corresponding to the IP [100] axis, which represents the magnetic hard axis of the film. In this configuration, the neutron spin is sensitive to the IP component of the magnetization, allowing the separation of the nuclear and magnetic SLD profiles. The applied field of 4.8 T corresponds to the maximum available at the instrument and remains well below the reported anisotropy field of SRO along the hard axis (12–14 T) \cite{koster2012structure}, implying that the film is not fully saturated in this direction. PNR measurements were performed at 100 K, 80 K, and 5 K, corresponding to representative temperatures selected based on the magnetometry results.

\begin{figure*}[h]
    \includegraphics[width=\linewidth]{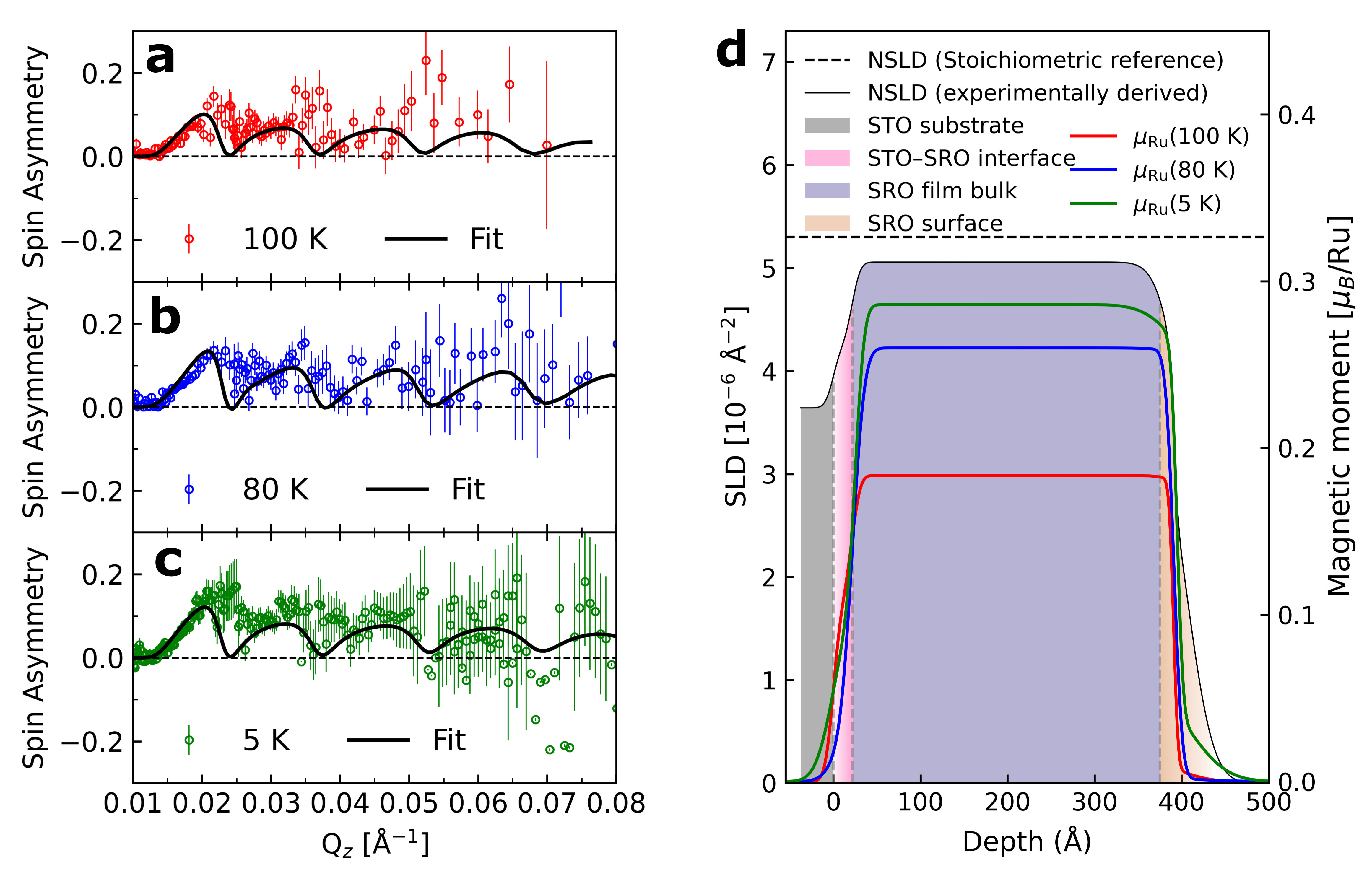}
    \caption{(a–c) Spin asymmetry [(R$+$) - (R$-$)]/[(R$+$) + (R$-$)] measured at (a) 100 K, (b) 80 K, and (c) 5 K after field cooling the sample in a 4.8 T in-plane magnetic field. Open symbols represent the experimental data and solid lines the fits obtained from the best-refined model. (d) Nuclear scattering length density (nSLD, left axis) and magnetization profiles (right axis) derived from the magnetic scattering length density (mSLD) extracted from the fits. The dashed line indicates the nominal nuclear SLD expected for stoichiometric bulk SrRuO$_3$. The reduced nSLD relative to the reference value indicates Ru deficiency extending throughout the film thickness. The interfacial regions at the film surface and at the SRO/STO interface, where the nSLD is further reduced and consistent with enhanced Ru deficiency, are highlighted by shaded areas. The magnetization profiles reveal finite magnetization at all temperatures, with reduced values in these interfacial regions compared to the film bulk.}
    \label{fig:PNR}
\end{figure*}

Figure \ref{fig:PNR}(a–c) shows the experimental spin asymmetry [(R+) - (R--)]/[(R+) + (R--)] curves together with the corresponding fits. The use of spin asymmetry emphasizes the magnetic contribution to the reflectivity by removing the purely structural background. The oscillatory behavior observed at 100 K, 80 K, and 5 K indicates a finite magnetic contrast within the film, which increases slightly upon cooling. Although the amplitude of the spin asymmetry is small, consistent with the reduced magnetic moment of Ru-deficient SRO, it is sufficient to extract reliable depth-dependent magnetization profiles. 

The corresponding nSLD and magnetization profiles obtained from the fits are shown in Fig. \ref{fig:PNR}(d). The magnetization profiles were derived from the mSLD obtained in the PNR fits, following the procedure described in the SM (Section \ref{section:sup_conversion}). The structural parameters extracted from PNR, including the film thickness and interface roughness, are in good agreement with those derived from XRR and AFM analyses, confirming the internal consistency of the structural model. Complementary to the XRR results, the PNR data indicate that Ru deficiency is not confined to the interfaces but extends throughout the entire film thickness, resulting in a uniformly reduced nSLD compared to the stoichiometric value [dashed line in Fig. \ref{fig:PNR}(d)]. A quantitative comparison between the nSLD obtained from PNR and the X-ray–derived SLD further supports this consistency (SM section \ref{section:sup_nsld}). Notably, the nSLD profiles exhibit further reductions in the interfacial regions, which appear as lighter shaded areas in Fig. \ref{fig:PNR}(d). These regions coincide with the interfaces identified by XRR and STEM-EDS and are therefore consistent with an enhanced Ru deficiency near both the film surface and the buried SRO/STO interface.

From the magnetic point of view, the film retains a finite magnetic signal across all temperatures, as shown by the magnetization profiles in Fig. \ref{fig:PNR}(d). Here, the magnetization profiles represent the in-plane component probed by PNR. The interface regions also exhibit measurable magnetic signals, although significantly reduced relative to the film bulk. For quantitative comparison with SQUID magnetometry, the depth-resolved magnetization profiles obtained from PNR were normalized to an effective magnetic moment per Ru atom ($\mu_B$/Ru), following the procedure described in the SM (section \ref{section:sup_conversion}). The average magnetic moment values extracted from PNR ($\langle\mu\rangle$ = 0.18, 0.26, and 0.29 $\mu_B$/Ru at 100, 80, and 5 K, respectively) were calculated as a thickness-weighted mean of the layer magnetizations according to $\langle\mu\rangle = \sum_i \mu_i t_i / \sum_i t_i$, where $t_i$ and $\mu_i$ are the thickness and magnetization of each layer, respectively. This procedure yields a thickness-weighted average in-plane magnetic moment per Ru, directly comparable to the SQUID values expressed in $\mu_B$/Ru.

The obtained values fall between the IP and OP SQUID magnetization components: at 100 K and 80 K, the PNR magnetization is closer to the OP SQUID values, while at 5 K it approaches the IP component. This evolution indicates that field cooling in 4.8 T induces a partial rotation of the magnetization vector toward the film plane, but complete IP alignment is not achieved. The small amplitude of the spin-asymmetry signal and the lack of full IP saturation confirm that a significant OP component persists, stabilized by the strong PMA. Therefore, the PNR results reveal a partially canted magnetic configuration, in which the magnetization vector gradually tilts toward the film plane upon cooling under an IP applied field. All SQUID and PNR measurements discussed here were performed following the same field-cooling protocol, and therefore, the observed evolution can be discussed in terms of intrinsic changes in the magnetic configuration under equivalent preparation conditions.

\begin{figure*}[hbtp]
    \makebox[\linewidth][c]{%
      \includegraphics[width=0.6\paperwidth]{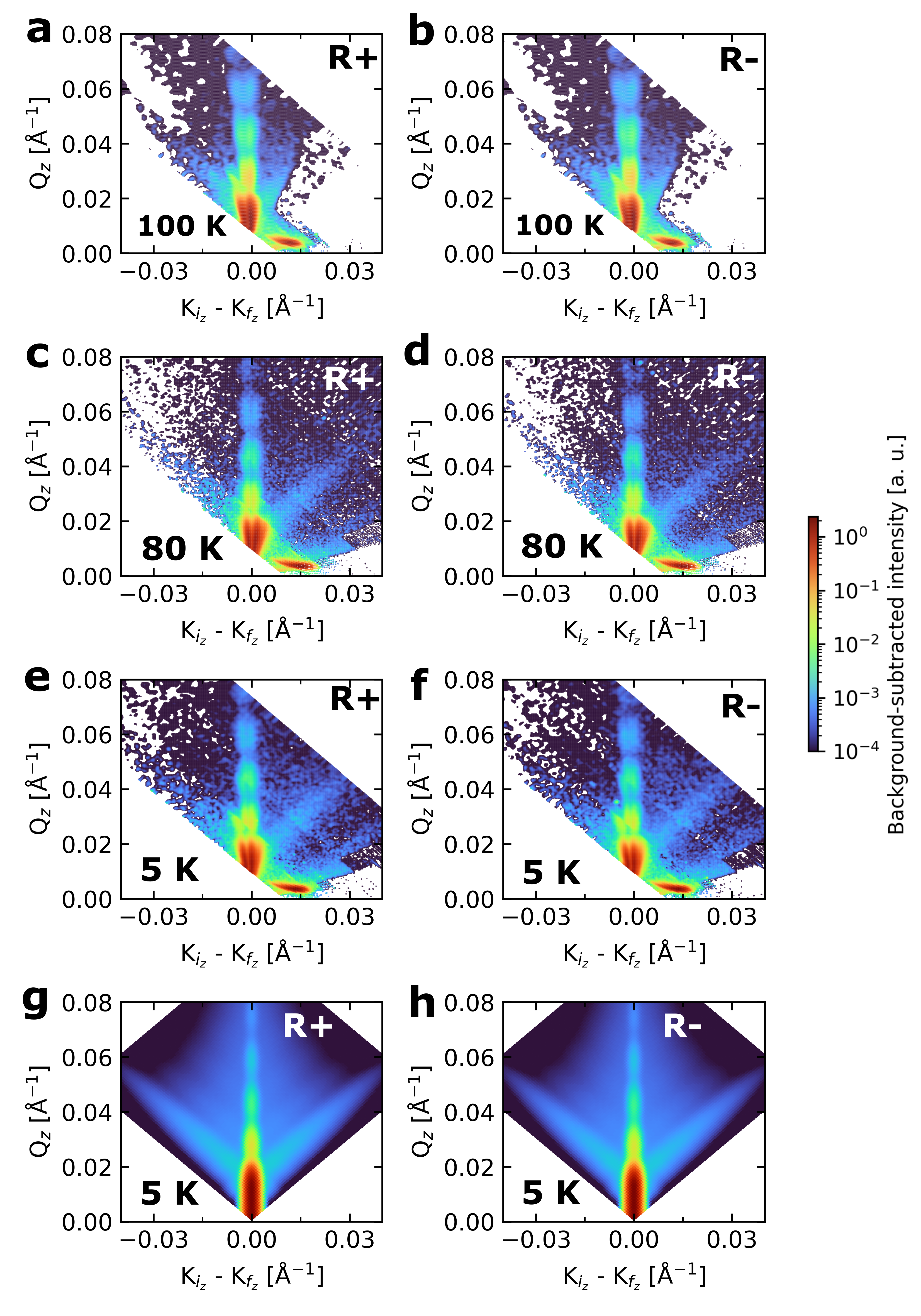}}
    \caption{Polarized off-specular neutron scattering (OSS) maps measured at 100 K, 80 K, and 5 K for both spin channels R+ (a,c,e) and R- (b,d,f), after field cooling in a 4.8 T in-plane magnetic field. The intensities were background-subtracted, and the same color scale is used for all temperatures to highlight relative changes in the scattering distribution. Panels (g,h) show simulated OSS patterns calculated from the PNR-derived structural and magnetic parameters at 5 K, reproducing the main experimental features.}
    \label{fig:OSS}
\end{figure*}

Figure \ref{fig:OSS} shows the OSS maps acquired at 100 K, 80 K, and 5 K for both spin channels (R$+$ and R$-$). The maps display the expected central specular line at $K_{i_z} - K_{f_z} = 0$ and diffuse scattering along the diagonal directions, characteristic of lateral structural or magnetic roughness and interfacial inhomogeneity, so called Yoneda scattering \cite{yoneda1963anomalous,lauter2007neutron,lauter20122}. The scattering is caused by the refraction and enhancement of the scattered waves at the surface and interface of a material, specifically when the angle of the scattered waves is close to the critical angle of the film. Upon cooling, an increase in the diffuse magnetic scattering intensity is observed. This increase is discussed in terms of a relative redistribution of the diffuse scattering intensity, as the maps are background-subtracted and compared using the same color scale. 

The OSS results provide complementary evidence supporting the depth-resolved magnetic profiles obtained from PNR. The enhanced diffuse scattering at low temperature is consistent with a gradual canting of the magnetization under the applied IP magnetic field. In this scenario, 
the strong PMA preserves a finite OP magnetization component, while the IP field promotes the development of a laterally inhomogeneous magnetic configuration with locally varying orientations. Such lateral magnetic inhomogeneity leads to a redistribution of magnetic correlations, resulting in enhanced OSS intensity. The OSS data therefore qualitatively support the PNR finding that the magnetization does not rotate uniformly into the film plane but instead evolves into a partially canted magnetic state at low temperature.

Simulated OSS maps [Figs. \ref{fig:OSS}(g,h)], performed accordingly Refs. \cite{kentzinger2008depth,ott2011off} using the structural and magnetic parameters extracted from the 5 K PNR fits, reproduce the main experimental features and the diagonal diffuse bands. This agreement indicates that the structural and magnetic roughness of the film and substrate are largely uncorrelated, consistent with island-like growth on TiO$_2$-terminated Nb:STO substrates. The analysis further suggests that, within the present experimental sensitivity, the lateral magnetic roughness closely follows the nuclear roughness profile, while the reduced magnetic moment of the Ru-deficient SRO film limits the overall magnetic contrast.

Given the substantial reduction in interfacial magnetization observed in Fig. \ref{fig:PNR}(d), it is plausible that the film interfaces may behave as magnetically inactive, or \textit{dead} layers. This interpretation is consistent with well-established observations in related complex oxide systems. For example, Porter et al. reported magnetic \textit{dead} layers of approximately 0.8–1.5 nm in La$_{0.7}$Sr$_{0.3}$MnO$_3$ (LSMO) films grown on STO substrates, depending on substrate type and film thickness \cite{porter2017magnetic}. Similar interfacial magnetic suppression has also been observed in SRO thin films. Horiuchi et al. recently described a $\sim$3 nm region near the SRO/STO interface as a \textit{dead} layer characterized by reduced magnetization and conductivity, although the microscopic origin of this behavior was not discussed in detail \cite{horiuchi2025single}. Other studies have also reported that near-interface regions in SRO exhibit diminished ferromagnetism or even non-magnetic behavior associated with disorder, stoichiometric deviations, or strain effects \cite{ali2022origin,falsetti2018high}. In manganite heterostructures, the emergence of such magnetically \textit{dead} layers has been linked to a combination of structural distortions, orbital reconstruction, oxygen vacancies, and interfacial strain \cite{pesquera2012surface,trappen2019depth,lima2021propriedades,schumacher2013inducing}. Similar mechanisms may also play a role in the interfacial magnetic suppression observed in ruthenate films.

In the present case, however, modeling the interfaces as completely non-magnetic did not yield satisfactory results (SM, section \ref{section:sup_pnr}). Such a model significantly altered the oscillation periodicity of the reflectivity curves, leading to unphysical estimates of the film thickness. These discrepancies indicate that, although the interfacial magnetization is strongly reduced, it does not vanish entirely. A more physically consistent explanation is that the PMA near the interfaces is stronger than in the film bulk, constraining the local magnetization to remain predominantly OP and thus reducing its IP projection detected by PNR. Accordingly, the refined model incorporating partially magnetized interfacial regions due to enhanced OP anisotropy reproduces the experimental data more accurately.

This finding highlights that Ru deficiency and local anisotropy variations at the interfaces play a central role in suppressing the IP magnetization, while the overall ferromagnetic integrity of the SRO film remains preserved. Importantly, these results show that the reduced interfacial magnetization often inferred in Ru-deficient SRO does not originate from magnetically \textit{dead} layers, but rather from strongly anisotropic interfacial magnetism.

\section{Conclusion}

In summary, a comprehensive set of complementary techniques demonstrates that SRO thin films grown by RF-HOPS are intrinsically Ru-deficient, with an enhanced deficiency at the interfaces and a finite deficiency extending throughout the film bulk. This non-stoichiometry strongly suppresses coherent electronic transport and reduces the saturation magnetization without significantly affecting the $T_{\text{Curie}}$, placing Ru-deficient SRO in a regime consistent with ferromagnetic insulator–like behavior. Depth- and lateral-resolved magnetic probes reveal that the reduced magnetization at the interfaces does not originate from magnetically \textit{dead} layers. Instead, the interfacial regions remain ferromagnetic and exhibit enhanced PMA, which constrains the local magnetization to remain predominantly out-of-plane and strongly reduces its in-plane projection. As a result, the magnetic response of Ru-deficient SRO films is governed by anisotropy gradients and partial canting rather than by a loss of magnetic order.

These findings highlight Ru deficiency as an effective control parameter for simultaneously tuning transport, magnetization, and anisotropy in correlated oxide thin films. These results demonstrate that defect and interface engineering provide effective routes to modulate interfacial magnetism beyond simple \textit{dead} layer scenarios, with direct relevance for the design of oxide heterostructures with tailored magnetic functionalities.

\section{Acknowledgments} 

The authors gratefully acknowledge the technical assistance of F. Gossen, L. Kibkalo and B. Schmitz. The authors acknowledge T. Denneulin, L. Jin, A. Kovács, O. Petracic and U. Poppe for the discussions. This work was sponsored by the
Tasso Springer Fellowship provided by JCNS. This research used resources at the Spallation Neutron Source, a DOE Office of Science User Facility operated by the Oak Ridge National Laboratory. The beam time was allocated to MAGREF on proposal number IPTS-32745.1. 

\section{Author contributions}

V.A.O.L. conceived the idea, designed the experiments, and led the project. V.A.O.L. grew the samples. V.A.O.L., M.I.F., O.C., and A.S. carried out the sample characterizations. V.L., A.Q., H.A., V.A.O.L., C.B.M., E.K., and S.N. designed the neutron scattering measurements and participated in the beamtime. V.A.O.L., A.Q., and V.L. analyzed the data. E.K. performed the off-specular neutron scattering simulations. All authors contributed to the discussion of the data. V.A.O.L. wrote the original manuscript with input from all authors. 

\bibliography{references}

\clearpage
\begin{center}
    \large{\textbf{Supplementary Material (SM) for\\
    \textit{Anisotropy-driven interfacial magnetism in Ru-deficient SrRuO$_3$ thin films}}}
\end{center}




\author{V. A. de Oliveira Lima}
\affiliation{Jülich Centre for Neutron Science for Quantum Materials and Collective Phenomena (JCNS-2), Forschungszentrum Jülich GmbH, 52425 Jülich, Germany}
\affiliation{RWTH Aachen, Faculty of Mathematics, Computer Science and Natural Sciences, 52074 Aachen, Germany}

\author{S. Nandi, T. Brückel}
\affiliation{Jülich Centre for Neutron Science for Quantum Materials and Collective Phenomena (JCNS-2), Forschungszentrum Jülich GmbH, 52425 Jülich, Germany}
\affiliation{RWTH Aachen, Faculty of Mathematics, Computer Science and Natural Sciences, 52074 Aachen, Germany}

\author{M. I. Faley}
\affiliation{Ernst Ruska-Centre for Microscopy and Spectroscopy with Electrons (ER-C-1), Forschungszentrum Jülich GmbH, 52425 Jülich, Germany}

\author{O. Concepción}
\affiliation{Peter Grünberg Institute for Semiconductor Nanoelectronics (PGI-9), Forschungszentrum Jülich GmbH, 52425 Jülich, Germany}

\author{A. Qdemat, H. Ambaye, V. Lauter}
\affiliation{Neutron Scattering Division, Oak Ridge National Laboratory, Oak Ridge, Tennessee  37831, United States of America}

\author{M. Radovic}
\affiliation{PSI Center for Photon Science, Paul Scherrer Institut, CH-5232 Villigen PSI, Switzerland}

\author{A. Singh, E. Kentzinger, C. Bednarski-Meinke}
\affiliation{Jülich Centre for Neutron Science for Quantum Materials and Collective Phenomena (JCNS-2), Forschungszentrum Jülich GmbH, 52425 Jülich, Germany}

\date{\today}
\maketitle

\renewcommand{\figurename}{SM FIG.}
\setcounter{figure}{0}
\renewcommand{\thefigure}{S\arabic{figure}}
\renewcommand{\theHfigure}{SM.\arabic{figure}}

\setcounter{section}{0}
\renewcommand{\thesection}{S\arabic{section}}
\renewcommand{\thesubsection}{S\arabic{section}.\arabic{subsection}}

\section{Models}
\label{section:sup_s1}

\begin{figure}[hbtp]
    \centering
    \includegraphics[width=\linewidth]{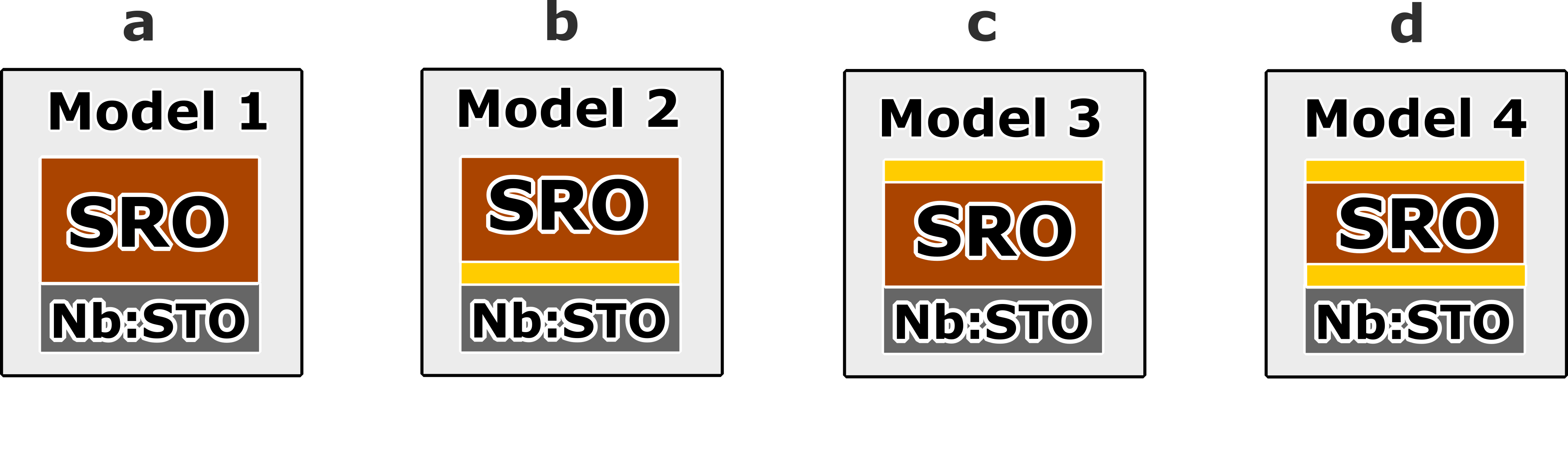}
    \caption{Schematic representation of the implemented models used to adjust the X-ray and Polarized Neutron reflectivity data. The yellow regions represent SRO/Nb:STO substrate and/or SRO surface/air interfaces.}
    \label{fig:sup1}
\end{figure}

Several block-like models were used to model the X-ray reflectivity (XRR) and polarized neutron reflectometry data. The schematic representation of the implemented models is shown in SM Fig. \ref{fig:sup1}. 

Model 1 [SM Fig. \ref{fig:sup1}(a)] treats the SRO film as a uniform single layer with no explicit interfaces. In Models 2 [SM Fig. \ref{fig:sup1}(b)] and 3 [SM Fig. \ref{fig:sup1}(c)], a thin SRO interface with reduced scattering length density (SLD) is introduced at the film-substrate and film–air interfaces, respectively. Model 3 assumes a coalesced film with a rougher surface arising from closely spaced islands, and provides a more plausible scenario considering the AFM micrographs presented in the main text. Model 4 [SM Fig. \ref{fig:sup1}(d)] includes reduced-SLD interfaces at both the top and bottom of the film, with the surface assumed to be rougher than the underlying layer. In the fits across all models, the top interface thickness was allowed to vary between 4 and 10 Å, while the bottom interface thickness was allowed to reach up to 30 Å, as previously reported in \cite{mlynarczyk2024physicochemical,mlynarczyk2007surface}.

\section{X-ray reflectivity fits}
\label{section:sup_s2}

\begin{figure*}[t]
    \centering
    \includegraphics[width=\linewidth]{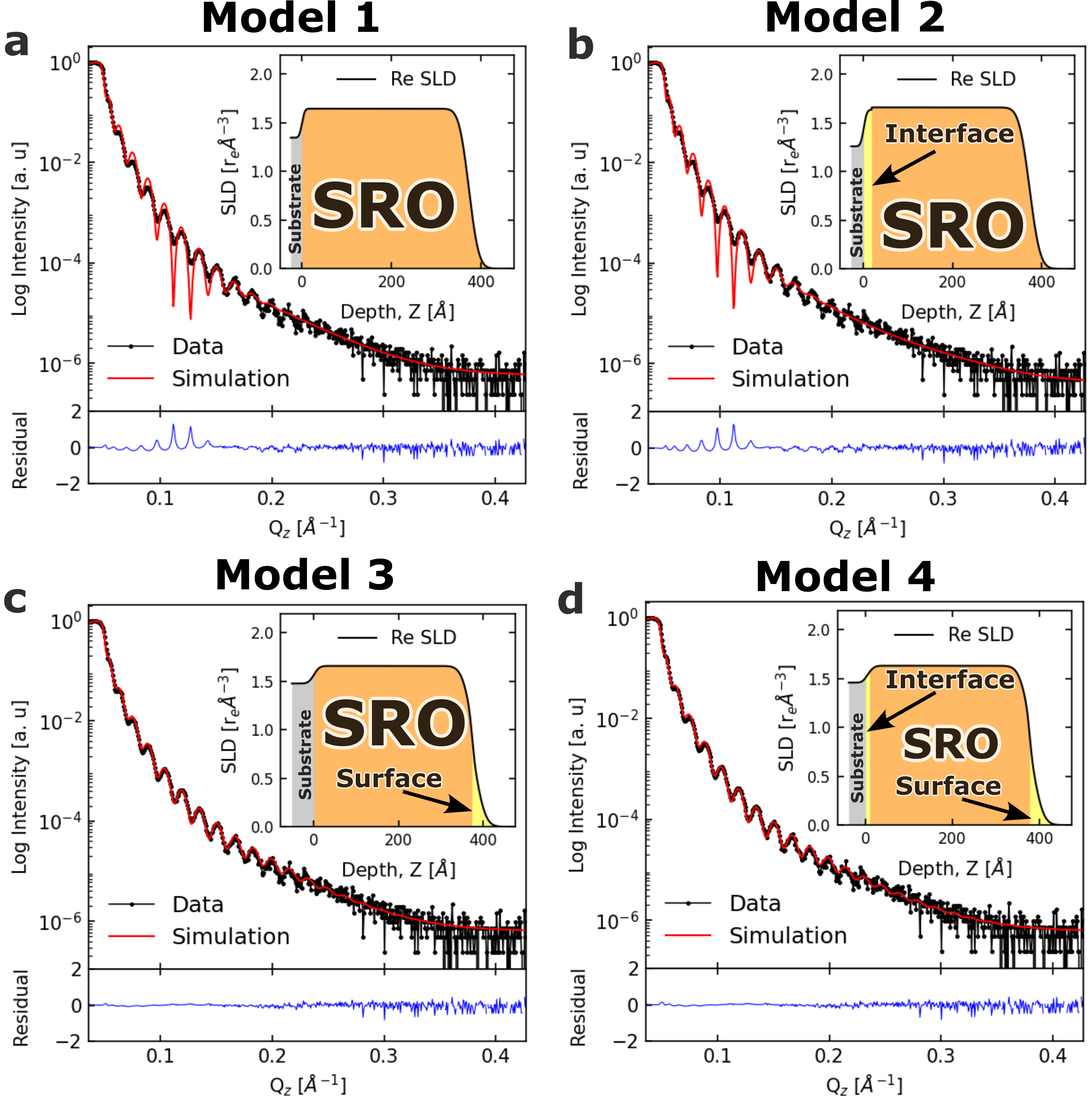}
    \caption{X-ray reflectivity (XRR) data (black symbols) and GenX simulations (red lines) for all models tested in this work. The insets show the real part of the SLD profiles used in each model. Residuals (blue lines) quantify the difference between experimental data and simulations.} 
    \label{fig:sup2}
\end{figure*}

\begin{figure}[hbtp]
    \centering
    \includegraphics[width=\linewidth]{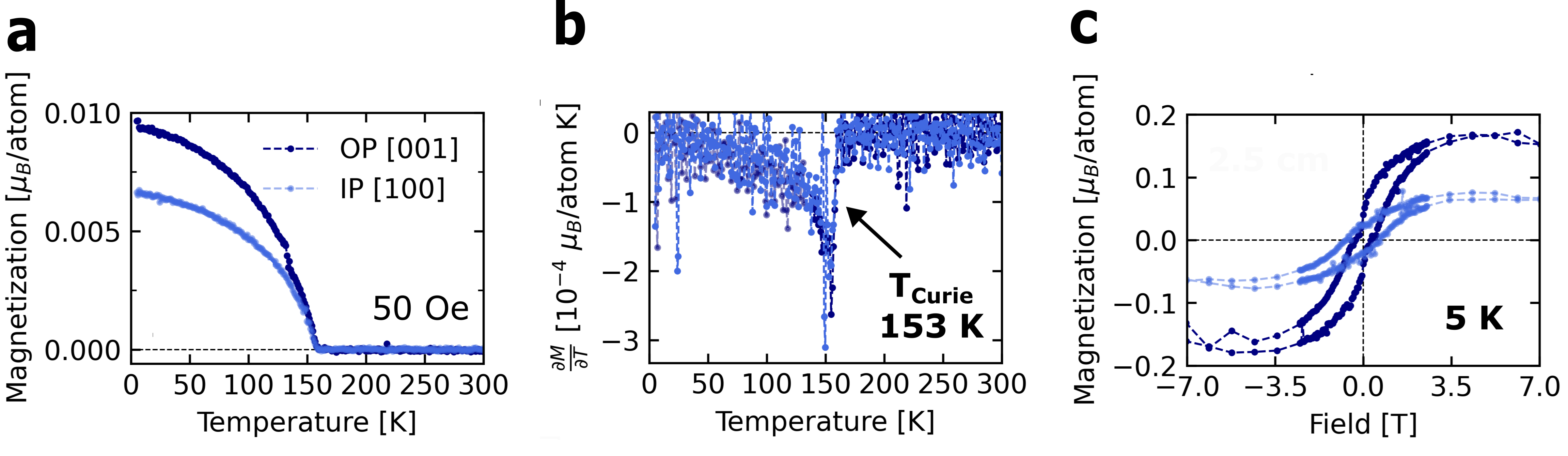}
    \caption{(a) Temperature-dependent magnetization $M$(T) of a Ru-deficient SRO thin film grown on undoped STO, measured after field cooling under an applied magnetic field of 50 Oe along the out-of-plane [001] and in-plane [100] directions. (b) The temperature derivative d$M$/dT, highlighting the $T_\text{Curie} \approx$ 153 K. (c) The corresponding magnetization hysteresis loops measured at 5 K.}
    \label{fig:sup3}
\end{figure}

The XRR data (black symbols), GenX fitting results (red lines), and residuals for all models tested are presented in SM Fig. \ref{fig:sup2}. The insets illustrate the corresponding SLD profiles obtained from the fits. The fitting results for all tested models are summarized in the main text and supporting figures. Detailed numerical parameters are available upon request.

Although Models 1 and 2 [SM Fig. \ref{fig:sup2}(a–b)] reproduce well the periodicity of the XRR oscillations and yield reasonably accurate film thicknesses $t$, they fail to adequately capture the interfacial roughness $\sigma$, particularly in the $\vb Q_\text{z}$ range of 0.07–0.15 Å$^{-1}$. In contrast, fits obtained using Models 3 [SM Fig. \ref{fig:sup2}(c–d)] and 4 [SM Fig. \ref{fig:sup2}(e)] provide an excellent overall agreement with the experimental data. Considering the local roughness observed in the AFM micrographs presented in the main text, Model 3 is likely more representative of the actual sample in comparison with Models 1 and 2. 

Model 4, similarly to Model 3, shows excellent agreement with the XRR data and is consistent with findings by M. Mlynarczyk et al. \cite{mlynarczyk2024physicochemical,mlynarczyk2007surface}, which report Sr surface and interface enrichment due to Ru volatility during SRO film growth. The FOM for Model 4 is nearly identical to that of Model 3, introducing some ambiguity as to which of the two models more accurately describes the sample. This was further investigated by STEM-EDS, as reported in the main text. However, taking in consideration the findings of M. Mlynarczyk et al. we are prone to agree that model 4 is the most appropriate. 

Looking at the residual plots (blue curves), a systematic offset is observed near the critical angle of total external reflection across all models. This may be related to the use of Nb-doped STO substrates, which, at the doping level considered in this study, exhibit lattice parameters, SLD and surface energies that are comparable to those of undoped STO, which may not be fully captured by the structural models employed in the fitting procedure.

Finally, we observed a reduction in the SLD not only at the top and/or bottom interfaces of the SRO film, but also throughout the entire film thickness. This suggests a generalized Ru deficiency in the SRO layer, with this deficiency being more pronounced at the interfaces. These results contrast with the findings of M. Mlynarczyk et al. \cite{mlynarczyk2024physicochemical,mlynarczyk2007surface}, who reported Sr enrichment only at the interfaces, while the film bulk remained stoichiometric. As reported in the main text, this generalized Ru deficiency affects the physical properties of SRO films, which usually show semiconductor-to-insulator electron transport and reduced magnetization. Interestingly, the magnetocrystalline perpendicular magnetic anisotropy and Curie temperature $T_{\text{Curie}}$ characteristic of SRO films are conserved \cite{oliveira2025anomalous}.

\clearpage
\section{Estimate of Ru deficiency from X-ray SLD}
\label{section:sup_S3}

In this section we estimate the effective Ru deficiency from the measured X-ray SLD obtained by XRR. The calculation follows a simple compositional model in which the film is described as SrRu$_x$O$_3$ (Ru vacancies only), while Sr and O contents are assumed fixed. X-ray SLD values are expressed in units of $r_e\,\text{\AA}^{-3}$, where $r_e$ is the classical electron radius.

\subsection{Atomic number density}
Assuming a pseudocubic perovskite unit cell with lattice parameter $a = 3.9~\text{\AA}$ and 5 atoms per formula unit (Sr + Ru + 3O), the atomic number density is
\begin{equation}
N = \frac{5}{a^3}
      = \frac{5}{(3.9~\text{\AA})^3}
      \approx 0.084~\text{\AA}^{-3}.
\end{equation}

\subsection{X-ray SLD of SrRu$_x$O$_3$}
For an X-ray energy corresponding to Cu K$\alpha$ radiation ($\lambda = 1.54~\text{\AA}$, $E = 12.3984/\lambda \approx 8.05~\text{keV}$), we write the X-ray SLD as
\begin{equation}
\rho_X(x) = N\, r_e\, \overline{(Z+f')}(x),
\end{equation}
where $\overline{(Z+f')}(x)$ is the average effective scattering factor per atom (including the dispersion correction $f'$ at $E=8.05$~keV), given by
\begin{equation}
\overline{(Z+f')}(x) =
\frac{1}{5}\Big[(Z+f')_{\mathrm{Sr}} + x\,(Z+f')_{\mathrm{Ru}} + 3\,(Z+f')_{\mathrm{O}}\Big].
\end{equation}
Considering
\begin{equation}
(Z+f')_{\mathrm{Sr}} = 37.6,\qquad
(Z+f')_{\mathrm{Ru}} = 44.1,\qquad
(Z+f')_{\mathrm{O}}  = 8.05,
\end{equation}
one obtains for stoichiometric SRO ($x=1$)
\begin{equation}
\overline{(Z+f')}(1)=\frac{1}{5}\left(37.6+44.1+3\times 8.05\right)=21.17,
\end{equation}
and therefore
\begin{equation}
\rho_X(1)=N\,r_e\,\overline{(Z+f')}(1)\approx 0.084\times 21.17
\approx 1.78~r_e\,\text{\AA}^{-3}.
\end{equation}

\subsection{Solving for the Ru occupancy $x$}
From XRR, the X-ray SLD in the film interior is measured as
\begin{equation}
\rho_X^{\mathrm{meas}} = 1.60~r_e\,\text{\AA}^{-3}.
\end{equation}

We solve $\rho_X(x)=\rho_X^{\mathrm{meas}}$ for $x$:
\begin{align}
\rho_X^{\mathrm{meas}}
&= N\,r_e \,\frac{1}{5}\Big[(Z+f')_{\mathrm{Sr}} + x\,(Z+f')_{\mathrm{Ru}} + 3\,(Z+f')_{\mathrm{O}}\Big],\\[4pt]
x
&=\frac{\displaystyle \frac{5\,\rho_X^{\mathrm{meas}}}{N}-(Z+f')_{\mathrm{Sr}}-3\,(Z+f')_{\mathrm{O}}}{(Z+f')_{\mathrm{Ru}}}.
\end{align}

Substituting $N=0.084~\text{\AA}^{-3}$ and the values above,
\begin{equation}
x=\frac{\displaystyle \frac{5\times 1.60}{0.084}-37.6-3\times 8.05}{44.1}
\approx 0.76.
\end{equation}

Thus, the effective Ru deficiency is
\begin{equation}
\delta_{\mathrm{Ru}} = 1-x \approx 0.24,
\end{equation}
i.e., an estimated Ru vacancy concentration of $\sim$24\% (SrRu$_{0.76}$O$_3$).

\clearpage
\section{Nuclear SLD}
\label{section:sup_nsld}

Using neutron coherent scattering lengths (in fm) $b_{\mathrm{Sr}}=7.02$, $b_{\mathrm{Ru}}=7.03$, and $b_{\mathrm{O}}=5.8$, the average coherent scattering length per atom for SrRu$_x$O$_3$ is
\begin{equation}
\overline{b}(x)=\frac{1}{5}\left(b_{\mathrm{Sr}}+x\,b_{\mathrm{Ru}}+3\,b_{\mathrm{O}}\right).
\end{equation}

The corresponding nuclear SLD is
\begin{equation}
\rho_N(x)=N\,\overline{b}(x),
\end{equation}
with $1~\mathrm{fm}=10^{-5}~\text{\AA}$. 

For stoichiometric SRO ($x=1$),
\begin{align}
\overline{b}(1)
&= \frac{1}{5}\left(7.02 + 7.03 + 3 \times 5.8\right)
= 6.29~\mathrm{fm}, \\
\rho_N(1)
&\approx 0.084 \times 6.29 \times 10^{-5}
\approx 5.3 \times 10^{-6}~\text{\AA}^{-2}.
\end{align}

For $x=0.76$ (24\% Ru deficiency),
\begin{align}
\overline{b}(0.76)
&= \frac{1}{5}\left(7.02 + 0.76 \times 7.03 + 3 \times 5.8\right)
= 5.97~\mathrm{fm}, \\
\rho_N(0.76)
&\approx 0.084 \times 5.97 \times 10^{-5}
\approx 5.0 \times 10^{-6}~\text{\AA}^{-2}.
\end{align}

The nuclear SLD values obtained from this estimate are consistent with the depth-averaged nSLD profiles extracted from the PNR fits in Fig.~4(d).

\clearpage
\section{Complementary magnetization of Ru-deficient SRO on undoped STO}
\label{section:sup_complement}

For comparison with the Nb:STO-based samples discussed in the main text, magnetization measurements were also performed on Ru-deficient SRO films grown on undoped TiO$_2$-terminated STO substrates. The film exhibits a slightly reduced Curie temperature ($T_\text{Curie} \approx$ 153 K) and a lower magnetic moment compared to the Nb:STO case, while maintaining a robust ferromagnetic transition. No clear signature of the antiferrodistortive transition is observed in this sample. These results indicate that substrate-related effects can modulate the strength of the magnetic anisotropy without altering its qualitative behavior.

\clearpage
\section{Conversion from magnetization and mSLD to magnetic moment per Ru atom}
\label{section:sup_conversion}

For consistency between SQUID magnetometry and polarized neutron reflectometry (PNR), the magnetic data are expressed in terms of an effective magnetic moment per Ru atom ($\mu_B$/Ru).

The SQUID measurements provide the magnetization M, defined as the magnetic moment per unit volume. The total magnetic moment of the film is therefore given by
\begin{equation}
\mu_{\text{tot}} = M \, V ,    
\end{equation}
where $V$ is the film volume.

The effective magnetic moment per Ru atom is obtained by normalizing the total moment by the number of Ru atoms in the film,
\begin{equation}
\mu_{\text{Ru}} = \frac{\mu_{\text{tot}}}{N_{\text{Ru}}}.    
\end{equation}

Assuming a pseudocubic unit cell volume $V_{\text{uc}}$ and an average Ru occupancy $f$, the number of Ru atoms is
$N_{\text{Ru}} = f \, \frac{V}{V_{\text{uc}}}$.

This yields
\begin{equation}
    \mu_{\text{Ru}} = \frac{M \, V_{\text{uc}}}{f},
\end{equation}
which is expressed in units of $\mu_B$/Ru. In the present work, an average Ru deficiency of approximately 25$\%$ ($f$ = 0.75) was used, as determined from XRR, RBS, and PNR analyses.

In the PNR analysis, the magnetic scattering length density (mSLD) is related to the magnetization profile via
\begin{equation}
\mathrm{mSLD}(z) = C \, M(z),    
\end{equation}
where $C = 2.91 \times 10^{-9}\,\text{Å}^{-2}\,\mu_B^{-1}$ is the neutron magnetic constant.

The magnetization profile $M(z)$ is converted to an effective magnetic moment per Ru atom using the same normalization procedure as described above, yielding depth-dependent values expressed in $\mu_B$/Ru.

This conversion uses an average Ru occupancy and therefore represents an effective magnetic moment per Ru atom, without resolving local variations of Ru deficiency at the interfaces.

\clearpage
\section{Polarized neutron reflectometry}
\label{section:sup_pnr}
Polarized Neutron Reflectometry (PNR) is a powerful technique for probing the structural and magnetic properties of thin films and buried interfaces. It provides depth-resolved information on both the nuclear and magnetic structure of materials, including layer-by-layer density, roughness, and magnetization profiles. The nuclear scattering length density (nSLD) characterizes the atomic structure, while the magnetic scattering length density (mSLD) reflects the in-plane magnetization as a function of depth. In off-specular neutron scattering (OSS), the magnetic signal arises from lateral fluctuations of the magnetization component that is perpendicular to the scattering wave vector $\Vec{Q} = Q_x\vec{e}_x + Q_z\vec{e}_z$ where $\vec{e}_x$ and $\vec{e}_z$ are unit vectors oriented parallel and perpendicular, respectively, to the film plane.

In the OSS geometry, the in-plane component ($Q_x$) of the scattering vector is much smaller than the perpendicular-to-plane
component ($Q_z$), i.e., $Q_x \ll Q_z$. As a result, OSS is primarily sensitive to lateral fluctuations of the in-plane component of the magnetization. The PNR and OSS measurements were performed with neutron spin polarization aligned parallel (R+) and antiparallel (R-) to the sample’s hard magnetization axis (in-plane) at various temperatures, after field cooling in a 5 T field applied parallel to the sample plane.  

\begin{figure*}
    \makebox[\linewidth][c]{%
      \includegraphics[width=0.9\paperwidth]{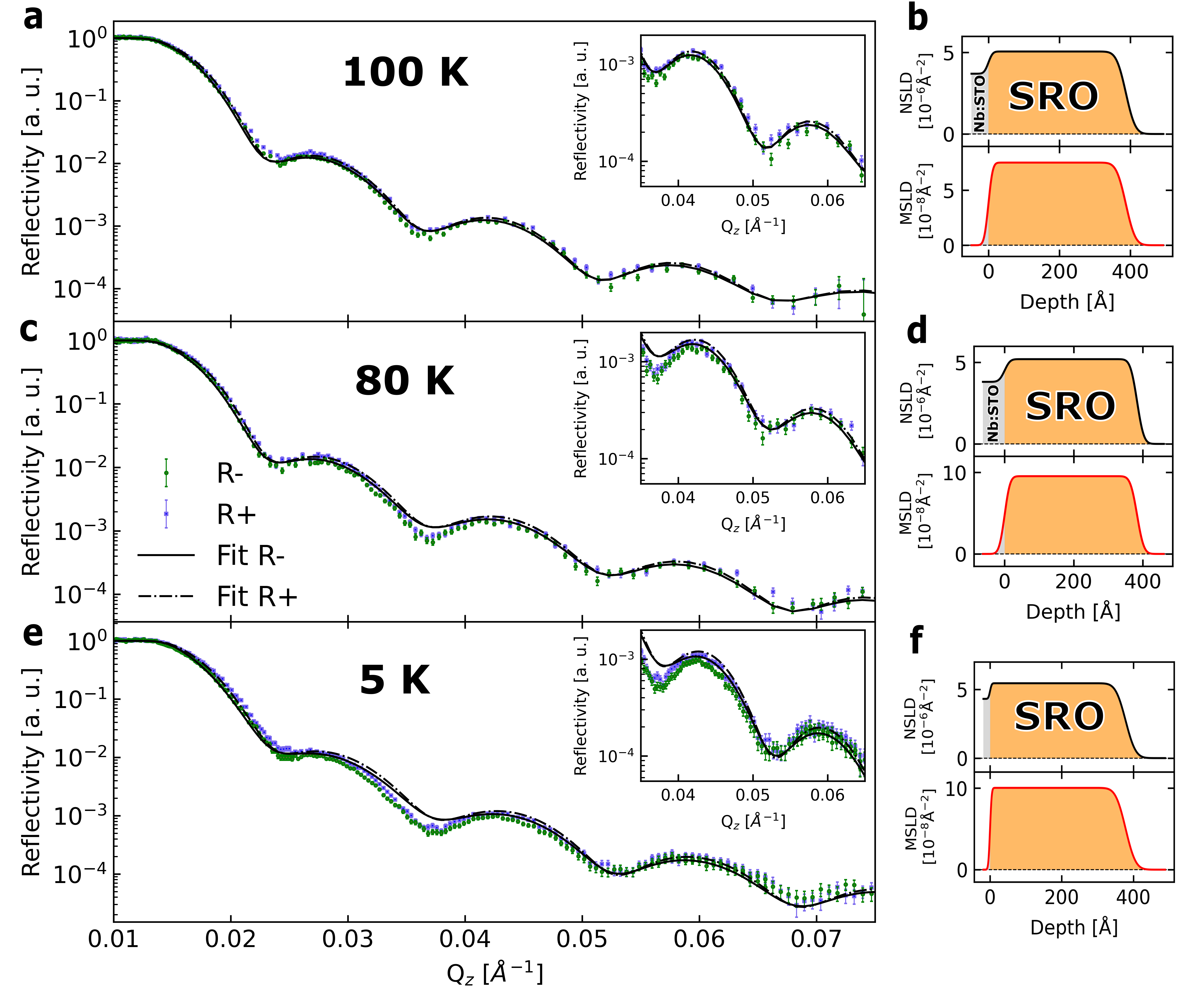}}
    \caption{Polarized neutron reflectometry (PNR) data acquired with neutron spin polarization parallel (R+) and antiparallel (R--) to the hard magnetization axis of the SRO thin film, along with the corresponding nuclear (nSLD) and magnetic (mSLD) scattering length density profiles at (a, b) 100 K, (c, d) 80 K, and (e, f) 5 K. The dotted-solid lines represent the fits to the R+ data, while the solid lines correspond to the R-- data, obtained accordingly to model 1. The insets offer a magnified view of the splitting between R+ and R-- curves as well as the fit quality.}
    \label{fig:sup4}
\end{figure*}

\begin{figure*}
    \makebox[\linewidth][c]{%
      \includegraphics[width=0.9\paperwidth]{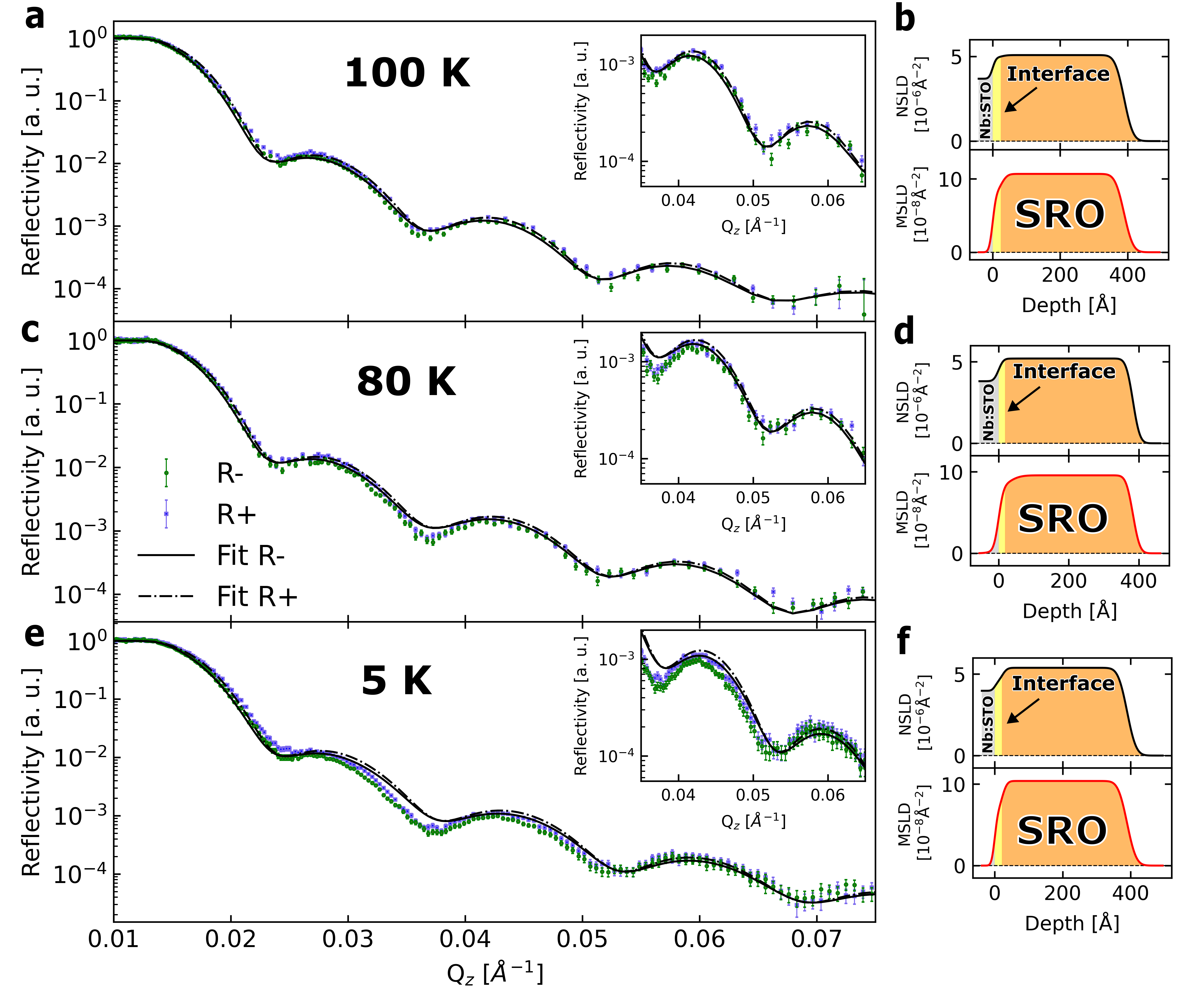}}
    \caption{Polarized neutron reflectometry (PNR) data acquired with neutron spin polarization parallel (R+) and antiparallel (R--) to the hard magnetization axis of the SRO thin film, along with the corresponding nuclear (nSLD) and magnetic (mSLD) scattering length density profiles at (a, b) 100 K, (c, d) 80 K, and (e, f) 5 K. The dotted-solid lines represent the fits to the R+ data, while the solid lines correspond to the R-- data, obtained accordingly to model 2. The insets offer a magnified view of the splitting between R+ and R-- curves as well as the fit quality.}
    \label{fig:sup5}
\end{figure*}

\begin{figure*}
    \makebox[\linewidth][c]{%
      \includegraphics[width=0.9\paperwidth]{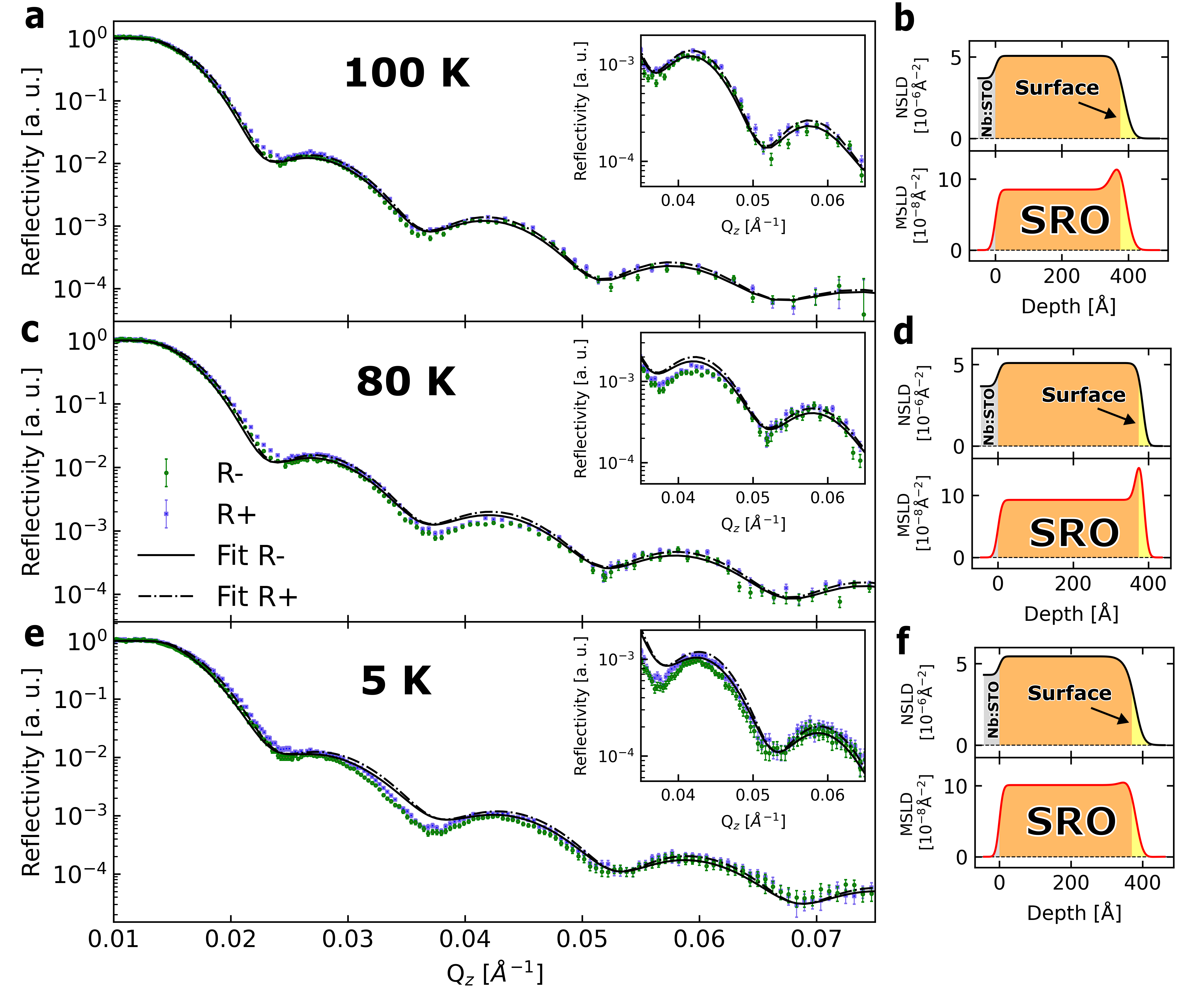}}
    \caption{Polarized neutron reflectometry (PNR) data acquired with neutron spin polarization parallel (R+) and antiparallel (R--) to the hard magnetization axis of the SRO thin film, along with the corresponding nuclear (nSLD) and magnetic (mSLD) scattering length density profiles at (a, b) 100 K, (c, d) 80 K, and (e, f) 5 K. The dotted-solid lines represent the fits to the R+ data, while the solid lines correspond to the R-- data, obtained accordingly to model 3-A. The insets offer a magnified view of the splitting between R+ and R-- curves as well as the fit quality.}
    \label{fig:sup6}
\end{figure*}

\begin{figure*}
    \makebox[\linewidth][c]{%
      \includegraphics[width=0.9\paperwidth]{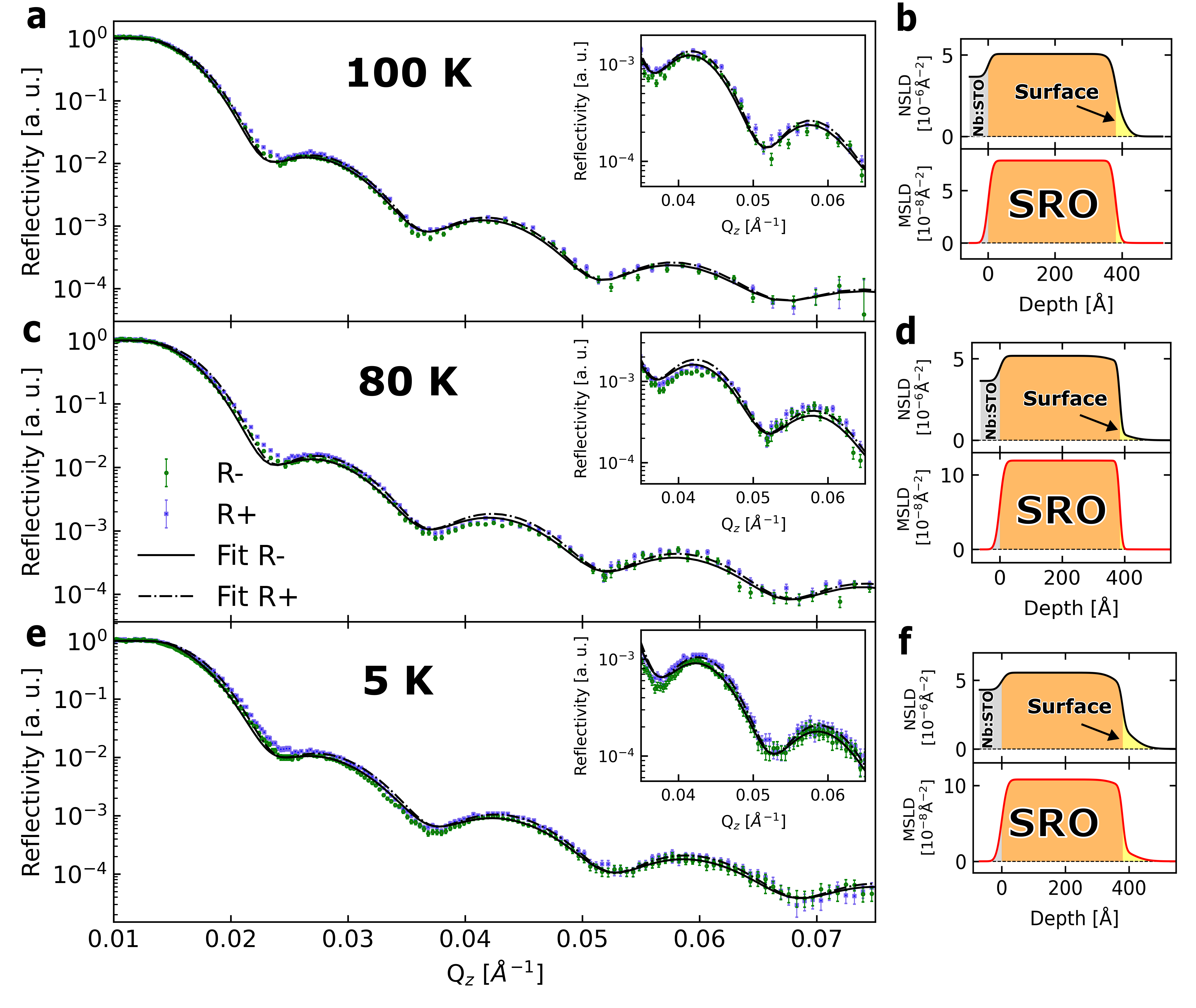}}
    \caption{Polarized neutron reflectometry (PNR) data acquired with neutron spin polarization parallel (R+) and antiparallel (R--) to the hard magnetization axis of the SRO thin film, along with the corresponding nuclear (nSLD) and magnetic (mSLD) scattering length density profiles at (a, b) 100 K, (c, d) 80 K, and (e, f) 5 K. The dotted-solid lines represent the fits to the R+ data, while the solid lines correspond to the R-- data, obtained accordingly to model 3-B. The insets offer a magnified view of the splitting between R+ and R-- curves as well as the fit quality.}
    \label{fig:sup7}
\end{figure*}

\begin{figure*}
    \makebox[\linewidth][c]{%
      \includegraphics[width=0.9\paperwidth]{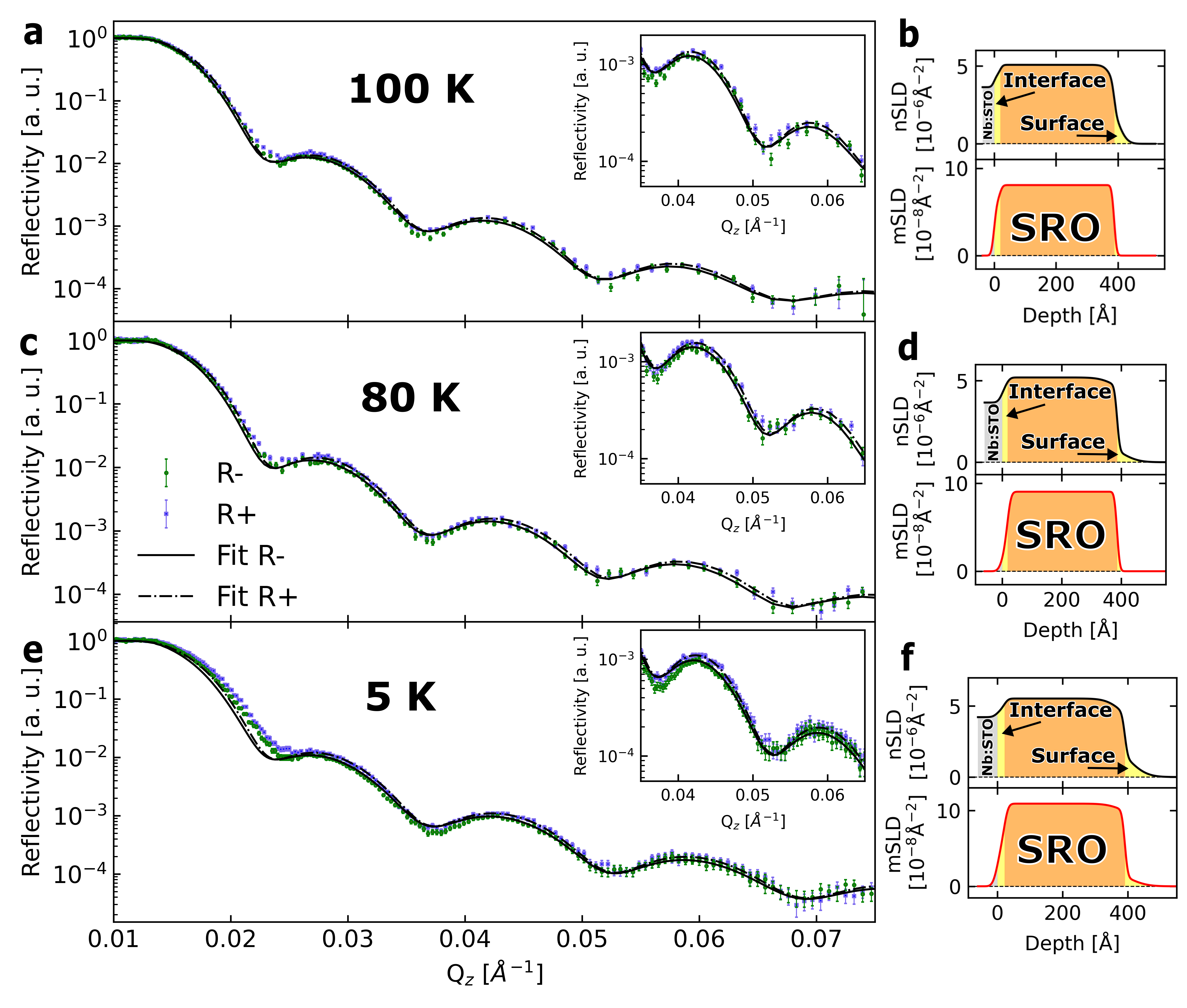}}
    \caption{Polarized neutron reflectometry (PNR) data acquired with neutron spin polarization parallel (R+) and antiparallel (R--) to the hard magnetization axis of the SRO thin film, along with the corresponding nuclear (nSLD) and magnetic (mSLD) scattering length density profiles at (a, b) 100 K, (c, d) 80 K, and (e, f) 5 K. The dotted-solid lines represent the fits to the R+ data, while the solid lines correspond to the R-- data, obtained accordingly to model 4-A (shown in the main text as spin-asymmetry). The insets offer a magnified view of the splitting between R+ and R-- curves as well as the fit quality.}
    \label{fig:sup8}
\end{figure*}

\begin{figure*}
    \makebox[\linewidth][c]{%
      \includegraphics[width=0.9\paperwidth]{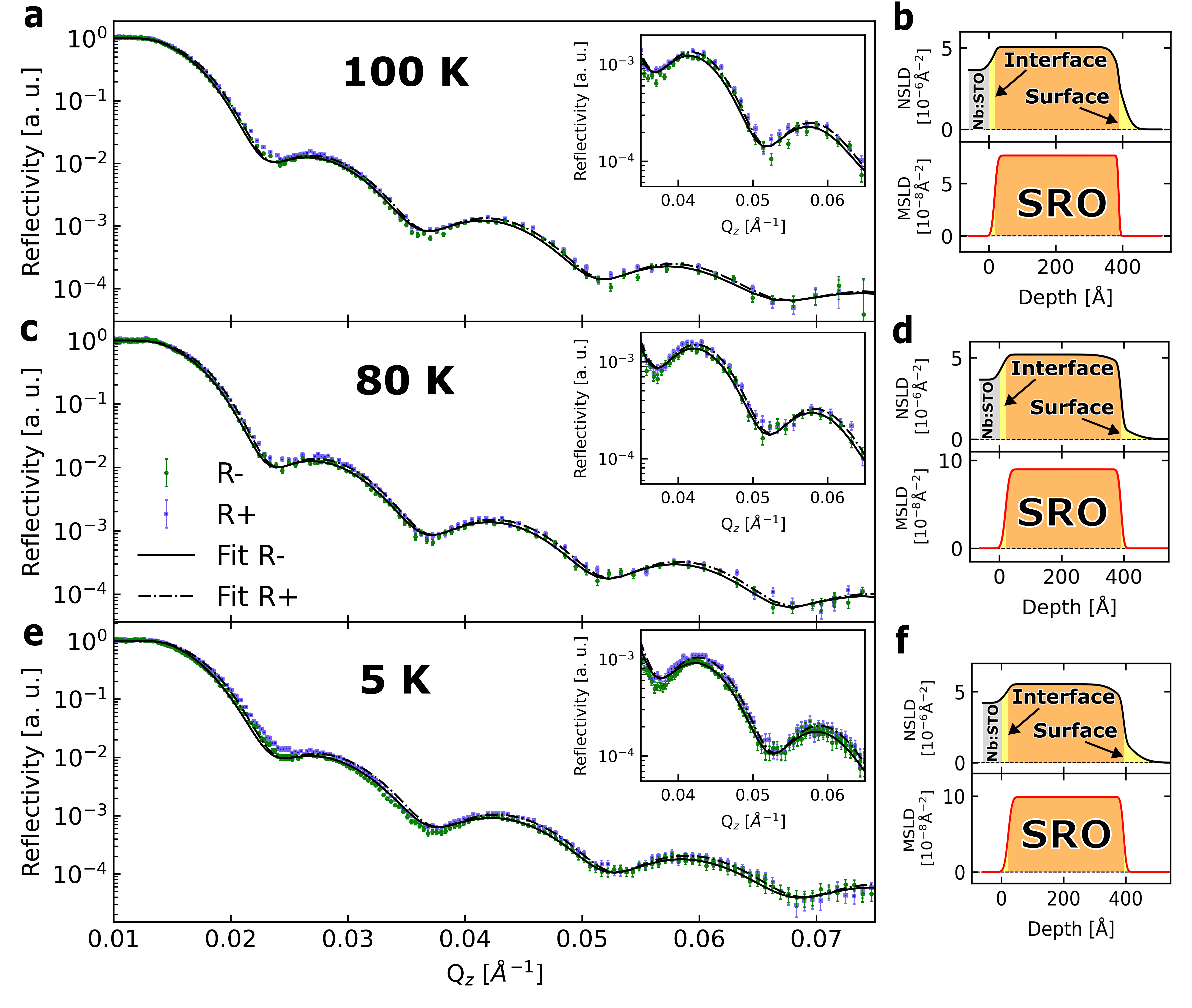}}
    \caption{Polarized neutron reflectometry (PNR) data acquired with neutron spin polarization parallel (R+) and antiparallel (R--) to the hard magnetization axis of the SRO thin film, along with the corresponding nuclear (nSLD) and magnetic (mSLD) scattering length density profiles at (a, b) 100 K, (c, d) 80 K, and (e, f) 5 K. The dotted-solid lines represent the fits to the R+ data, while the solid lines correspond to the R-- data, obtained accordingly to model 4-B. The insets offer a magnified view of the splitting between R+ and R-- curves as well as the fit quality.}
    \label{fig:sup9}
\end{figure*}

Given the small splitting between the R+ and R- PNR curves, it can be directly inferred that the SRO film exhibits a reduced magnetization. Nevertheless, this does not prevent a detailed investigation of how the magnetic properties of the sample are affected by the Ru deficiency. In this context, the PNR data were analyzed using models analogous to those employed in the XRR analysis, now incorporating the sample’s magnetic moment information. SM Figures \ref{fig:sup4}--\ref{fig:sup9} show the fitting results obtained for each structural model and its magnetic variants. The PNR experimental data, fits, and residuals are shown for three temperatures: (a) 100 K, (b) 80 K, and (c) 5 K. In each panel, the upper inset displays the corresponding nuclear scattering length density (nSLD) and the magnetic scattering length density (mSLD) profiles extracted from the fit. The fitting results for all tested models are summarized in the main text and supporting figures. Detailed numerical parameters are available upon request.

As observed in the XRR fitting, models 1 and 2 also fail to adequately describe the PNR data. Although they reasonably capture the overall film thickness, they do not account for interfacial roughness, which appears to play a critical role. At 100 K, the apparent agreement between the models and the data may be misleading, as the limited $\vb{Q}_\text{z}$ range (up to 0.08 Å$^{-1}$) reduces sensitivity to narrow or diffuse interface regions. The mismatch between the fits and the experimental data becomes more evident at lower temperatures and higher $\vb{Q}_\text{z}$ ranges, where interfacial features are more clearly resolved. 

For the PNR analysis, we have modified the Model 3 to explore two possible magnetic surface scenarios, from now on called Model 3-A and Model 3-B. While the first considers a rough film morphology with widely spaced islands that could act as nucleation centers for magnetic domains (magnetic pinning) at low temperatures, as suggested by low temperature magnetic force microscopy (Low T-MFM) studies on SRO thin films grown on low-miscut STO substrates \cite{zahradnik2020magnetic, wang2020magnetic}, the latter assumes a coalesced film with a simply rougher surface arising from closely spaced islands without any magnetic pinning possibility. In both cases the surface layer was considered Ru deficient and, therefore, have reduced nSLD values. 

Model 3-A yields an intriguing mSLD profile characterized by a pronounced magnetic enhancement at the film surface, suggesting a possible magnetic pinning by the islands. However, this strong magnetization near the surface appears physically questionable, since Ru is the primary magnetic ion in SRO and is known to be deficient in both surface and interface regions under the high oxygen pressure sputtering (HOPS) growth conditions. However, recent studies have shown that surface-localized magnetic moments or pinned magnetic regions can emerge even in Ru-deficient areas, provided that the local atomic structure, lattice distortions, or electronic inhomogeneities break the symmetry and induce magnetic frustration. In this context, surface magnetism may arise less from stoichiometric composition (e.g., Sr enrichment or Ru deficiency) and more from structural effects such as disorder (e.g. islands). For instance, Palai et al. reported spin-glass-like behavior in ultrathin SRO films with extremely low roughness, which they attributed to magnetic frustration at interfaces and surfaces rather than to compositional variations \cite{palai2009observation}. 

These findings support the possibility that rough surface features like islands could host pinned moments or short-range correlated spins, even in the absence of bulk-like ferromagnetism. Nevertheless, this sounds out of context considering that we have no clear evidence of structural disorder or magnetic frustration in our samples prepared by radio-frequency HOPS, and therefore, the physical plausibility of model 3-A is strongly limited. Although the model reproduces a magnetic surface enhancement at the surface, that can be characteristic of a possible magnetic pinning, it fails to account for key experimental features, such as match the magnetization values obtained by comparative techniques (SQUID magnetometry). These inconsistencies suggest that, despite the model fits well the features of the data, it lacks physical justification under the current growth conditions. 

Models 3-B and 4-A (shown in the main text as spin-asymmetry) provide a good representation of the PNR data at all measured temperatures, consistent with the trends observed in the XRR results. The extracted values of film thickness $t$ and interfacial roughness $\sigma$ are in good agreement with those obtained from deposition rate calibrations, XRR, and AFM measurements. In addition, the magnetization values (in units of $\mu_B$/atom) extracted from both models are very similar to the ones obtained by SQUID magnetometry. This consistency, however, reinforces a recurring ambiguity that could only be resolved by directly probing the Ru concentration profile at the film interfaces with the STEM-EDS. Based on the STEM-EDS results and the available evidence, particularly the results reported by M. Młynarczyk et al. \cite{mlynarczyk2024physicochemical, mlynarczyk2007surface}, we considered model 4-A as the most physically justified between these two. 

Given the substantial reduction in the interfacial magnetic moment extracted using model 4‑A, it is plausible that the film interfaces may be effectively non-magnetic. The interfacial regions at both the top and bottom of the film are ultrathin, and the magnetism within these layers may be extremely weak, as indicated by model 4‑A, or may correspond to magnetically dead layers. To test this hypothesis, model 4‑B was evaluated by imposing zero magnetic moment in both interfacial regions. This approach is consistent with well-established observations in related complex oxide systems. 

For example, Porter et al. (2019) studied La$_{0.7}$Sr$_{0.3}$MnO$_3$ (LSMO) films grown on STO substrates and reported magnetic dead layers ranging from approximately 0.8 to 1.5 nm, depending on substrate type and film thickness \cite{porter2017magnetic}. These dead layers exhibited strongly suppressed ferromagnetism at the interfaces, despite being structurally continuous. Similarly, other studies on manganite–oxide heterostructures have linked the emergence of magnetic dead layers to mechanisms such as structural distortions, orbital reconstruction, oxygen vacancies, and interfacial strain \cite{pesquera2012surface,trappen2019depth,lima2021propriedades}. Considering this analogy, it is plausible that a similar effect may occur in our sample: ultrathin interface regions might exhibit significantly reduced or even vanishing magnetization, despite structural continuity. However, in our case, modeling the interfaces as fully magnetically dead did not yield satisfactory results since it significantly altered the oscillation periodicity in the reflectivity curves, leading to unrealistic estimates for the film thickness.

As a final remark, we noted that the first minimum in the PNR oscillation pattern is not well fitted by either model. This mismatch is attributed to the instrumental resolution and the data extraction and merging process applied during PNR reduction.

\end{document}